\documentclass[aps,preprint,twocolumn,superscriptaddress,10pt]{revtex4}

\usepackage{amsfonts}
\usepackage{amsmath}
\usepackage{amssymb}
\usepackage{graphicx}
\usepackage{dcolumn}
\usepackage{mciteplus}
\usepackage{euscript}
\usepackage{multirow}
\usepackage{color}

\providecommand{\U}[1]{\protect\rule{.1in}{.1in}}
\newcommand{\redtx}[1]{\textcolor{red}{}}
\newcommand{\rfeL}{$R$Fe$_{12}$}
\newcommand{\rfeti}{$R$Fe$_{11}$Ti}

\newcommand{\rcoti}{$R$Co$_{11}$Ti}
\newcommand{\rfecoti}{$R$(Fe$_{1-x}$Co$_{x}$)$_{11}$Ti}
\newcommand{\yfeL}{YFe$_{12}$}

\newcommand{\yfeti}{YFe$_{11}$Ti}
\newcommand{\yfetih}{YFe$_{11}$TiH}
\newcommand{\yfetin}{YFe$_{11}$TiN}
\newcommand{\yfetic}{YFe$_{11}$TiC}
\newcommand{\ycoL}{YCo$_{12}$}

\newcommand{\ycoti}{YCo$_{11}$Ti}
\newcommand{\ycotih}{YCo$_{11}$TiH}
\newcommand{\ycotic}{YCo$_{11}$TiC}
\newcommand{\ycotin}{YCo$_{11}$TiN}
\newcommand{\yfecoti}{Y(Fe$_{1-x}$Co$_{x}$)$_{11}$Ti}
\newcommand{\yfecotiz}{Y(Fe$_{1-x}$Co$_{x}$)$_{11}$Ti$Z$}
\newcommand{\cefeL}{CeFe$_{12}$}

\newcommand{\cefeti}{CeFe$_{11}$Ti}
\newcommand{\cefetih}{CeFe$_{11}$TiH}
\newcommand{\cefetin}{CeFe$_{11}$TiN}
\newcommand{\cefetic}{CeFe$_{11}$TiC}
\newcommand{\cecoL}{CeCo$_{12}$}

\newcommand{\cecoti}{CeCo$_{11}$Ti}
\newcommand{\cecotih}{CeCo$_{11}$TiH}
\newcommand{\cecotin}{CeCo$_{11}$TiN}
\newcommand{\cecotic}{CeCo$_{11}$TiC}

\newcommand{\cefecoti}{Ce(Fe$_{1-x}$Co$_{x}$)$_{11}$Ti}
\newcommand{\cefecotiz}{Ce(Fe$_{1-x}$Co$_{x}$)$_{11}$Ti$Z$}
\newcommand{\ksoc}{$K_\text{so}$}
\newcommand{\kso}{$K_\text{so}$}

\newcommand{\morb}{$M_{l}$}

\def\mspin{$m_\text{s}$}
\def\morb{$m_l$}

\newcommand{\mevfuB}{($\frac{\text{meV}}{\text{f.u.}}$)}
\newcommand{\mjmcB}{($\frac{\text{MJ}}{\text{m}^3}$)}
\newcommand{\mubfuB}{($\frac{\mu_B}{\text{f.u.}}$)}

\newcommand{\mubfu}{\,$\mu_B$/f.u.}
\newcommand{\mevat}{\,meV$/$atom}
\newcommand{\mjmc}{\,MJm$^{-3}$}
\newcommand{\KK}{\,K}

\def\mub{$\mu_B$}

\def\tc{{$T_\text{C}$}}
\def\dtc{{$\Delta T_\text{C}$}}
\def\dm{{$\Delta M$}}
\def\dk{{$\Delta K$}}

\def\etal{$\textit{et al.}$}
\def\abinito{$\emph{ab initio}$}

\newcommand{\fk}[1]{\footnotemark[#1]}
\newcommand{\mc}[3]{\multicolumn{#1}{#2}{#3}}
\newcommand{\mr}[3]{\multirow{#1}{#2}{#3}}

\newcommand{\req}[1]{Eq.~(\ref{#1})}

\newcommand{\rfig}[1]{Fig.~\ref{#1}}
\newcommand{\rtbl}[1]{Table~\ref{#1}}

\begin{document}
\title{Intrinsic magnetic properties of
  {$R$(Fe$_{1-x}$Co$_{x}$)$_{11}$Ti$Z$} ($R$ = Y and Ce; $Z$ = H, C,
  and N)}

\author{Liqin Ke}
\email[Corresponding author: ]{liqinke@ameslab.gov}
\affiliation{Ames Laboratory US Department of Energy, Ames, Iowa 50011}
\author{Duane D. Johnson}
\affiliation{Ames Laboratory US Department of Energy, Ames, Iowa 50011}
\affiliation{Materials Science $\&$ Engineering, Iowa State University, Ames, Iowa 50011-2300}
\begin{abstract}
To guide improved properties coincident with reduction of critical
materials in permanent magnets, we investigate via density functional
theory (DFT) the intrinsic magnetic properties of a promising system,
{\rfecoti}$Z$ with $R$=Y, Ce and interstitial doping ($Z$=H, C, N).
The magnetization $M$, Curie temperature {\tc}, and magnetocrystalline
anisotropy energy $K$ calculated in local density approximation to DFT
agree well with measurements.  Site-resolved contributions to $K$
reveal that all three Fe sublattices promote uniaxial anisotropy in
{\yfeti}, while competing anisotropy contributions exist in {\ycoti}.
As observed in experiments on {\rfecoti}, we find a complex
nonmonotonic dependence of $K$ on Co content, and show that anisotropy
variations are a collective effect of MAE contributions from all sites
and cannot be solely explained by preferential site occupancy.  With
interstitial doping, calculated {\tc} enhancements are in the sequence
of N$>$C$>$H, with volume and chemical effects contributing to the
enhancement.  The uniaxial anisotropy of {\rfecoti}$Z$ generally
decreases with C and N; although, for $R$=Ce, C doping is found to
greatly enhance it for a small range of 0.7$<$$x$$<$0.9.
\end{abstract}

\eid{identifier}
\date{\today}
\maketitle

\section{Introduction}
 
\redtx{Why {\cefeL} and why doping?}
The search for new permanent magnets without critical materials has
generated great interest in the magnetism
community.\cite{mccallum.arms2014,kusne.srep2014} Developing
{\cefeL}-based rare-earth($R$)-transition-metal($TM$)
intermetallics\cite{yang.jap1988,buschow.jmmm1991,pan.jap1994,sakuma.jpsj1992,korner.sr2016}
is an important approach, considering the relative abundance of Ce
among $R$ elements and the large content of inexpensive Fe. To improve
{\cefeti} as a permanent magnet, it is desired to modify the compound
to achieve the best possible intrinsic magnetic properties, such as
magnetization $M$, Curie temperature {\tc}, and magnetocrystalline
anisotropy energy (MAE) $K$. Both substitutional doping with
Co\cite{yang.ssc1988} and interstitial doping with small elements of
H, C or N can strongly affect its magnetic properties. A theoretical
understanding of intrinsic magnetic properties in this system and the
effect of doping will help guide the experiments and help ascertain
the best achievable permanent magnet properties.

\redtx{Compound formation, why Ti and why Y?} 
Binary iron compounds of $R$Fe$_{12}$ do not form for any $R$ elements
unless a small amount of stabilizer elements are added, such as
$T$=Ti, Si, V, Cr, Mo, or W.\cite{hmm.v07c5} Such $R$Fe$_{12-z}T_{z}$
compounds are generally regarded as ternaries rather than
pseudobinaries because the third element, $T$, atoms often have a very
strong site preference and exclusively sit at one of three
nonequivalent Fe sites.\cite{hmm.v06c1} Magnetization often decreases
quickly with the increase of $T$ composition and a minimum amount of
Ti ($z$=0.7) is needed to stabilize the structure, resulting in Ti
compounds having better magnetic properties than
others.\cite{hu.jap1990} Prototype yttrium compounds are often studied
to focus on the properties of the $TM$ sublattices in the
corresponding $R$-$TM$ systems because yttrium can be regarded as a
nonmagnetic, rare-earth element.

\redtx{M, J, $T_c$ and K in pure compound:} 
In comparison to other $R$-Fe systems,\cite{hu.jap1990,coey.jmmm1989}
such as Y$_2$Fe$_{17}$ and Y$_2$Fe$_{14}$B, Fe sublattices in 1-12
compounds have relative low magnetization due to a more compact
structure, but at low temperatures a very high uniaxial MAE, e.g.,
$K$=$2${\mjmc}in YFe$_{11}$Ti.\cite{yang.ssc1988,deboer.jlcm1987}
Curie temperatures are relatively low; $M$ and $K$ quickly decrease
with increasing
temperature.\cite{solzi.jap1988,isnard.jac1998,tereshina.jpcm2001}
{\cefeti} has {\tc}$\approx$485{\KK}, and a low-temperature
magnetization within a range of $17.4-20.2$ {\mubfu}, while {\yfeti}
has a slightly larger $M$ and {\tc}. At room temperature, {\cefeti}
has a larger $K$ (1.3{\mjmc}) than YFe$_{11}$Ti (0.89{\mjmc}). This
may indicate that the Ce sublattice has a positive contribution to the
uniaxial anisotropy.\cite{isnard.jac1998}

\redtx{Substitutional doping with Co:} 
The substitutional doping with Co is a common approach to improve
{\tc} in $R$-Fe compounds.\cite{yang.ssc1988} Pure phase {\rfecoti}
exists over the whole composition range for both $R$=Y and
Ce.\cite{zhou.jap2014} The largest magnetization in {\yfecoti} occurs
at YFe$_8$Co$_3$Ti while the {\tc} increases continuously with Co
composition until it reaches the maximum in
{\ycoti}.\cite{ohashi.jap1991,wang.jpcm2001} For Ce compounds, the
maximum {\tc} is obtained in CeFe$_{2}$Co$_{9}$Ti.\cite{zhou.jap2014}
The dependence of MAE on the Co composition in {\rfecoti} is more
intriguing and not understood. Although early
experiments\cite{yang.ssc1988,solzi.jap1988} suggested that {\ycoti}
has a planar anisotropy, later experiments agreed that {\ycoti} has
uniaxial
anisotropy,\cite{sinha.jmmm1989,cheng.jap1991,ohashi.jap1991,moze.jpcm1995,wang.jmsj1999,wang.jpcm2001,zhou.jap2014}
but with a magnitude smaller than those of {\yfeti}. For the
intermediate Co composition, anisotropy changes from uniaxial to
planar and then back to uniaxial with the increase of Co composition
in both of {\yfecoti} and
{\cefecoti}.\cite{sinha.jmmm1989,cheng.jap1991,ohashi.jap1991,moze.jpcm1995,wang.jmsj1999,wang.jpcm2001,zhou.jap2014}

\redtx{Interstitial doping with H, C, and N:} 
The interstitial doping with H,\cite{isnard.jac1998} N
\cite{yang.apl1991,yang.ssc1991,pan.jap1994} and C
\cite{hurley.jpcm1992,qi.jpcm1992,li.jpcm1993} can increase $M$ and
     {\tc}, and provide control of the magnitude and sign of the MAE
     constants in {\rfeti}. Hydrogenation simultaneously increases all
     three intrinsic magnetic properties in {\yfeti}, and enhancements
     are {\dm}=1{\mubfu} at 4.2{\KK},
     \dtc=60{\KK},\cite{nikitin.ijhe1999,tereshina.jpcm2001} and
            {\dk}= 6.5$\%$,\cite{tereshina.jpcm2001}
            respectively. Insertion of larger C and N atoms has a much
            stronger effect on the enhancements of $M$ and
            {\tc}.\cite{qi.jpcm1992,nikitin.jac2001} Unfortunately, it
            is achieved at the expense of uniaxial anisotropy. In
            comparison with {\yfeti}, enhancements of
            {\dm}=2.6{\mubfu} and {\dtc}=154~K were observed in
            {\yfeti}C$_{0.9}$, and {\dm}=2.7{\mubfu} and
            {\dtc}=218{\KK} in {\yfeti}N$_{0.8}$. The MAE decreases by
            {\dk}=0.6$\sim$0.7 {\mjmc} in both compounds.  Doping
            influences the MAE contributions from both $TM$ sublattice
            and rare-earth atoms. It had been argued that the
            proximity of doping atoms to the rare-earth atoms in
            $R$TiFe$_{11}$N$_x$ may lead to drastic changes in the
            rare-earth sublattice anisotropy,\cite{yang.ssc1991} and N
            doping often has an opposite effect on MAE as H
            doping.\cite{nikitin.jac2001} For Ce compounds, a similar
            amount of {\tc} enhancement was obtained upon
            nitriding,\cite{pan.jap1994} and the effect of H doping is
            much smaller. Isnard {\etal}\cite{isnard.jac1998} found
            that not much change is observed upon H insertion either
            in the room temperature anisotropy or in saturation
            magnetization.

Other possible interstitial doping elements such as B, Si or P atoms
are much less favored to occupy the interstitial sites due to chemical
or structural reasons.\cite{hurley.jpcm1992} In fact, interestingly,
it has been found that the B atoms prefer to substitute for some of
the Ti atoms and drive the Ti into the
interstitial.\cite{dan.jpcm1995}

\redtx{Complication of Ce states and justification the plain LDA method?}
The nature of the Ce $4f$ state different Ce-$TM$ compounds is often a
controversial subject.\cite{coey.jap1993} The anomalies in the lattice
constants as well as the magnetic moment and Curie temperature have
been interpreted as evidence of the mixed-valence (between Ce$^{3+}$
and Ce$^{4+}$) behavior of the cerium ion. It is further complicated
by the doping. Controversy remains on how Ce valence states are
affected upon hydrogenation.\cite{chaboy.prb1995,isnard.jac1998} It
also has been shown that Ce $4f$ states are itinerant and, as such,
the standard localized $4f$ picture is not appropriate for systems
such as CeCo$_5$.\cite{nordstrom.prb1990,eriksson.prl1988} Moreover,
in the (Nd-Ce)$_2$Fe$_{14}$B system, the mixed valency of Ce has been
shown to be due to local site volume and site chemistry
effects.\cite{alam.apl2013} In this paper the $4f$ states in Ce are
treated as itinerant and included as valence states, and we found that
magnetic properties calculated are in good agreement with experiments.

\section{Calculation details}
\label{sec:2}

\subsection{Crystal structure}
\begin{figure}[tbp]
\includegraphics[width=.95\linewidth,clip,angle=0]{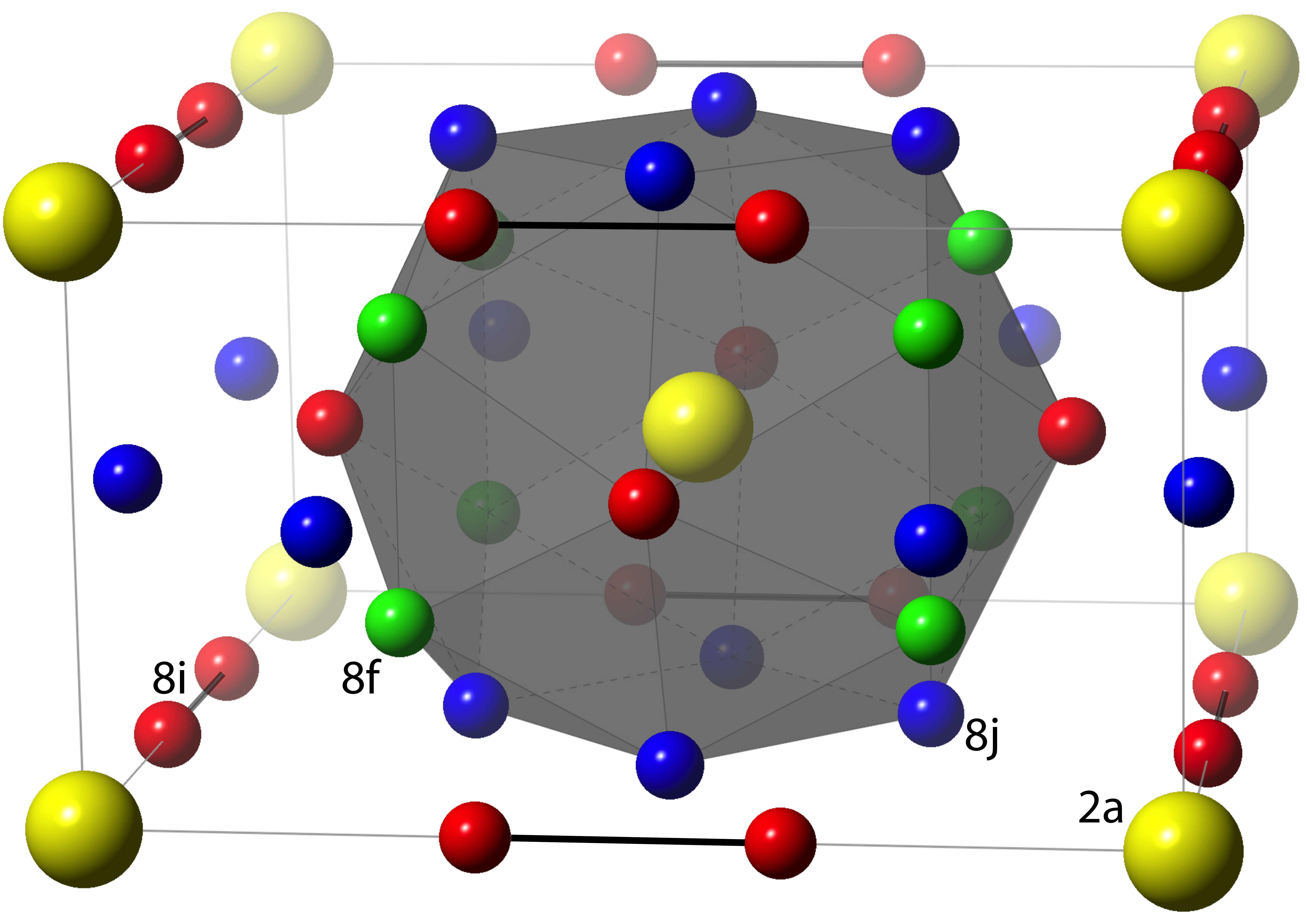}
\caption{ (Color online) Schematic representation of the crystal
  structures of {\rfeti}. $R(2a)$ atoms are indicated with larger
  spheres in yellow color. Three transition metal sublattices $8i$,
  $8j$, and $8f$ are in red, blue and green, respectively. Each $R$
  atom has four nearest $8i$, eight nearest $8j$, and $8f$ neighbor
  atoms. Among three $TM$ sites, the $8i$ site has the shortest
  distance from the $R$ atom. Interstitial sites $2b$ (not shown) are
  halfway between two $R$ atoms along the $c$ axis and coordinated by
  an octahedron of two $R$ and four Fe(8$j$) sites.}
\label{fig:xtal}
\end{figure}

{\rfeti} has a body-center-tetragonal ThMn$_{12}$-type ($I4/mmm$ space
group, no.139) structure, which is closely related to the 1-5 and 2-17
$R$-$TM$ structures.\cite{hu.jap1990} The primitive unit cell contains
one formula unit (f.u.). As shown in \rfig{fig:xtal}, $R$ atoms occupy
the $2a (4/mmm)$ site, while transition metal atoms are divided into
three sublattices, $8i$($m2m$), $8j$($m2m$) and $8f$($2/m$), each of
which has fourfold multiplicities. The $8j$ and $8f$ sites bear a
great similarity in their local environments with respect to the
distribution of coordinated atoms,\cite{wang.jac2001} whereas the $8i$
sites, often referred to as dumbbell sites, form -Fe-Fe-$R$- chains
with $R$ atoms along the basal axes, instead of the $c$ axis, as in
the 2-17 structure.\cite{ke.intermag2015} Ti atoms occupy nearly
exclusively on the $8i$ sites, however, the distribution of Fe and Ti
atoms within the 8$i$ sites is disordered.

To calculate {\rfeti}, we replace one of four Fe($8i$) atoms with Ti
in the primitive cell of {\rfeL} and neglect the effect of the
artificial Ti ordering introduced by using this unit cell. Although
the $I4/mmm$ symmetry is lowered by Ti substitution or the spin-orbit
coupling (SOC) in the anisotropy calculation, we still use the
notations of $8j$, $8i$, and $8f$ sites for simplicity.

For Co doping, M\"ossbauer spectroscopy found that Co atoms
preferentially occupy the $8f$ sites in {\yfecoti},\cite{li.jap1991}
while the high-resolution neutron powder diffraction experiments
concluded that Co atoms preferentially occupy sites in the sequence of
$8j>8f>8i$.\cite{liang.jap1999} For interstitial H, C, and N doping,
neutron scattering has shown that dopants prefer to occupy the larger
octahedral $2b$ interstitial sites,\cite{yang.ssc1991,isnard.jac1998}
which have the shortest distance from the rare-earth sites among all
empty interstitial sites. In all our calculations, we also assume that
H, C, or N atom occupies the $2b$ sites.

\subsection{Computational methods}
Most magnetic properties were calculated using a standard linear
muffin-tin orbital (LMTO) basis set\cite{andersen.prb1975} generalized
to full potentials.\cite{methfessel.chap2000} This scheme employs
generalized Hankel functions as the envelope functions. For MAE
calculation, the SOC was included through the force
theorem.\cite{mackintosh1980} The MAE is defined below as
$K$=$E_{110}-E_{001}$, where $E_{001}$ and $E_{110}$ are the summation
of band energies for the magnetization being oriented along the
$[001]$ and $[110]$ directions, respectively.  Positive (negative) $K$
corresponds to uniaxial (planar) anisotropy. It should be noted that,
due to the presence of Ti in the primitive cell, the two basal axes
become inequivalent, with -Ti-Fe-$R$- chains along the $[100]$
direction and -Fe-Fe-$R$- chains along the $[010]$
direction. $E_{100}$ and $E_{010}$ become different, which is an
artifact introduced by using the small primitive cell and artificial
ordering of Ti within the 8$i$ sublattice.

We found that the $[100]$ direction is harder than the $[010]$ in
{\yfeti}, and vice versa in {\ycoti}. $E_{110}$ is usually about the
average of $E_{100}$ and $E_{010}$. Thus, we use $[110]$ as the
reference direction for the basal plane.  A $16\times 16\times 16$
$k$-point mesh is used for MAE calculations to ensure sufficient
convergence; MAE in {\yfeti} changed by less than 3$\%$ when a denser
$32\times 32\times 32$ mesh was employed.  To decompose the MAE, we
evaluate the on-site SOC matrix element $\langle V_\text{so} \rangle$
and the corresponding anisotropy {\kso}=$\frac{1}{2}\langle
V_\text{so} \rangle_{110}{-}\frac{1}{2}\langle V_\text{so}
\rangle_{001}$.  Unlike MAE, {\ksoc} can be easily decomposed into
sites, spins, and orbital pairs. According to second-order
perturbation theory,\cite{antropov.ssc2014,ke.prb2015}
\begin{equation}
\label{eq:eq1} 
  K\approx\sum_{i}K_\text{so}(i),
\end{equation}
where $i$ indicates atomic sites. Equation~(\ref{eq:eq1}) holds true
for all compounds that we investigated in this paper.  Hence, we use
{\ksoc}($i$) to represent the site-resolved MAE. For simplicity, we
write it as $K(i)$.
 
Exchange coupling parameters $J_{ij}$ are calculated using a static
linear-response approach implemented in a Green's function (GF) LMTO
method, simplified using the atomic sphere approximation (ASA) to the
potential and density.\cite{ke.prb2013,ke.prb2012} The
scalar-relativistic Hamiltonian was used so SOC is not included,
although it is a small perturbation on $J_{ij}$'s. In the basis set,
$s, p, d, f$ orbitals are included for Ce, Y, Fe, and Co atoms, and
$s, p$ orbitals are included for H, C, and N atoms. Exchange
parameters $J_{ij}({\bf q})$ are calculated using a $16^3$ $k$-point
mesh, and $J_{ij}({\bf R})$ can be obtained by a subsequent
Fourier-transforming. {\tc} is estimated in the mean-field
approximation (MFA) or random-phase approximation (RPA). See
Ref.~\onlinecite{ke.prb2013} for details of the methods to calculate
{\tc}.

For all magnetic property calculations, the effective one-electron
potential was obtained within the local density approximation (LDA) to
DFT using the parametrization of von Barth and Hedin.\cite{barth}
However, with the functional of Perdew, Becke, and Ernzerhof (PBE)
being better at structural relaxation for most of the solids
containing $3d$ elements,\cite{haas.prb2009} we use it to fully relax
the lattice constants and internal atomic positions in a fast
plane-wave method, as implemented within the Vienna {\abinito}
simulation package (VASP).\cite{kresse.prb1993,kresse.prb1996} The
nuclei and core electrons were described by the projector augmented
wave (PAW) potential\cite{kresse.prb1999} and the wave functions of
valence electrons were expanded in a plane-wave basis set with a
cutoff energy of up to 520 $eV$. All relaxed structures are then verified in
FP-LMTO before the magnetic property calculations are performed.

\section{Results and discussion}
\label{sec:3}

\subsection{structure}

\begin{table}[tbp]
\caption{Calculated and measured (Exp.) values for the lattice
  parameters and volume are listed for various compounds.}
\label{tbl:qm}
\def\arraystretch{1.2}
\begin{tabular}{lcccccr}
\hline \hline
Compounds       &  a\footnotemark[1](\AA) & c(\AA) & V(\AA$^3$) & $\Delta V/V$ & Ref. &\\ \hline
{\yfeti}~(Exp.) &  8.480   & 4.771  &   343.08   &  &  [\onlinecite{obbade.jac1997}]   &   \\   
{\yfeti}        &  8.472   & 4.720  &   338.78   &   0        &  &   \\   
{\yfeL}         &  8.447   & 4.695  &   334.94   &  -1.1      &  &   \\   
{\yfetih}       &  8.457   & 4.732  &   338.43   &  -0.10     &  &   \\   
{\yfetic}       &  8.517   & 4.834  &   350.67   &   3.51     &  &   \\   
{\yfetin}       &  8.563   & 4.791  &   351.31   &   3.70     &  &   \\   
\hline          
{\ycoti}~(Exp.) &  8.367   & 4.712  & 329.87  &  &  [\onlinecite{moze.ssc1988}]   &        \\   
{\ycoti}        &  8.328   & 4.673  &  324.08    &  0       &  &      \\   
{\ycoL}         &  8.268   & 4.655  &  318.21    & -1.81    &  &      \\   
{\ycotih}       &  8.343   & 4.688  &  326.30    &  0.68    &  &      \\   
{\ycotic}       &  8.396   & 4.767  &  336.08    &  3.70    &  &      \\   
{\ycotin}       &  8.436   & 4.716  &  335.59    &  3.55    &  &      \\   
\hline
{\cefeti}~(Exp.) &  8.539   & 4.780  &  348.53    &  & [\onlinecite{isnard.jac1998}]   &  \\   
{\cefeti}       &  8.524   & 4.670  &  339.35    &  0      &  & \\   
{\cefeL}        &  8.504   & 4.648  &  336.12    & -0.95   &  & \\   
{\cefetih}      &  8.498   & 4.738  &  342.13    &  0.82   &  & \\   
{\cefetic}      &  8.501   & 4.891  &  353.45    &  4.16   &  & \\   
{\cefetin}      &  8.570   & 4.809  &  353.17    &  4.07   &  & \\   
\hline          
{\cecoti}~(Exp.)&  8.380   & 4.724  & 331.74     &  & [\onlinecite{zhou.jap2014}]     &   \\   
{\cecoti}       &  8.360   & 4.657  & 325.46     & 0      &  & \\   
{\cecoL}        &  8.291   & 4.648  &  319.51    & -1.82   &  &   \\   
{\cecotih}      &  8.359   & 4.694  & 327.94     & 0.76   &  &  \\   
{\cecotic}      &  8.383   & 4.811  & 338.07     & 3.87   &  &  \\   
{\cecotin}      &  8.442   & 4.735  & 337.44     & 3.68   &  &  \\  \hline  
 \hline
\end{tabular}\\ 
\footnotetext[1]{ Except for the hypothetical 1-12 compounds, Ti
  substitution in the 13-atom cell breaks the symmetry of {\cefeL},
  and lattice parameters $a$ and $b$ become nonequivalent. The listed
  calculated $a$ is an average of $a$ and $b$ of the unit cell used in
  the calculation.}
\end{table}

Lattice constants and volumes are listed in \rtbl{tbl:qm}, the
calculated lattice constants are in good agreement with experiments.
The strong Ti site preference on the $8i$
site\cite{yang.jap1988,moze.ssc1988,isnard.jac1998} had been
interpreted in terms of atomic volume, coordination number, and
enthalpy. It had been argued that enthalpy associated with $R$ and Ti,
V, or Mo atoms are positive and $8i$ sites have the smallest contact
area with $R$ atoms.  To identify quantitatively the site-preference
effect, we calculated the total energy of {\cefeti} with one Ti atom
occupying at the 8$i$, 8$j$, or 8$f$ sites, respectively, in the
13-atom primitive cell. The lowest-energy structure is the one with Ti
atoms on the $8i$ site. Energies are higher by 42 {\mevat} and 60
{\mevat} with Ti atom being on the {$8j$} and $8f$ sites,
respectively. Hence, Ti atom should have a strong preference to occupy
the $8i$ sites, as observed in the experiments.

In comparison to the hypothetical 1-12 compounds, the replacement of
Fe or Co atoms with Ti increases volume by $1\%$ or $2\%$,
respectively.  Experimentally, H doping slightly increases the volume
by 1\% in {\yfetih}, which is not observed in our calculation. The
calculated volume of {\cefetih} is 0.82\% larger than
{\cefeti}. Calculations show that carbonizing and nitriding have a
larger effect on volume expansion than hydrogenation and volume
expansion is larger in Ce compounds than in Y compounds, both of which
agree with experiments.

\begin{figure}[tbp]
\begin{tabular}{cc}
\includegraphics[width=.49\textwidth,clip,angle=0]{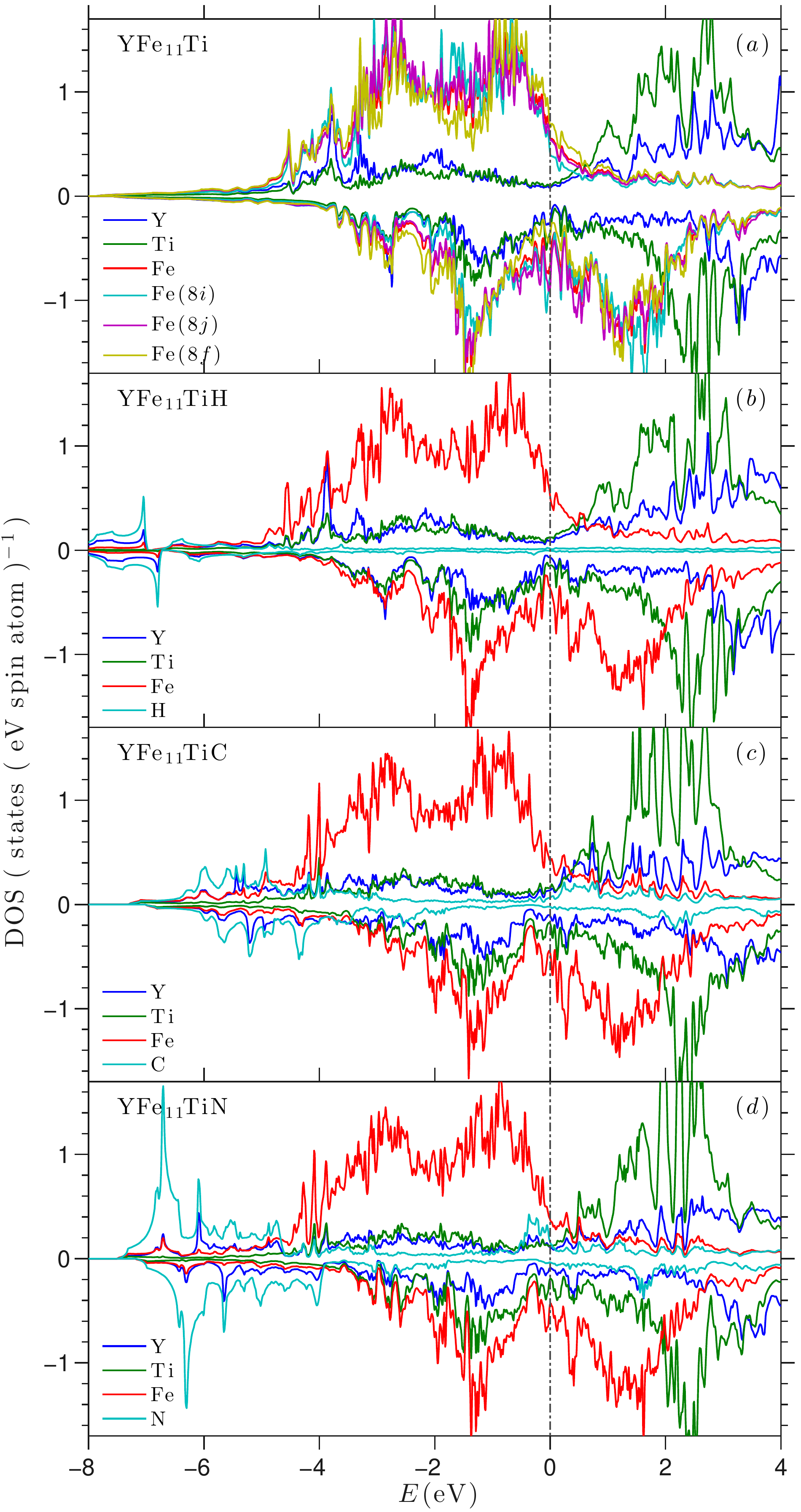} \\
\end{tabular}
\caption{ (Color online) Atom- and spin-projected, partial densities
  of states (DOS) in ($a$) {\yfeti}, ($b$) {\yfetih}, ($c$) {\yfetic},
  and ($d$) {\yfetin} within the LDA and no SOC.  For \yfeti, the of
  Fe DOS are further resolved by averaging states projected on $8i$,
  $8j$, and $8f$ sites. Majority spin (positive values) and minority
  spin (negative values) DOS are shown separately. Fermi energy $E_F$
  is at 0 eV.}
\label{fig:pdos}
\end{figure}

The total density of states of {\yfeti} and {\yfetin} compares
reasonably well with previously reported LMTO-ASA
calculations.\cite{sakuma.jpsj1992} Figure.~\ref{fig:pdos} shows the
scalar-relativistic partial density of states (PDOS) projected on
individual elements in {\yfeti}, {\yfetih}, {\yfetic}, and
{\yfetin}. The Fe PDOS are averaged over 11 Fe atoms. The interstitial
doping elements on $2b$ sites hybridizes with neighboring $R$ and
Fe($8j$) atoms. H-$s$ states hybridizes with neighboring Y and
Fe($8j$) atoms at around -7 eV below the Fermi level in {\yfetih}. The
C-$p$ and N-$p$ states have larger energy dispersion in {\yfetic} and
{\yfetin}, respectively. The Fe states hybridized with interstitial
elements, as shown in \rfig{fig:pdos}, are mostly from four (8$j$) out
of 11 Fe sites. Fe($8f$) sites are the furthest away from the
interstitial $2b$ sites and their hybridization with doping elements
are negligible. The Ce compounds have a large $f$-states above the
Fermi level and share lots of similar PDOS features with the
corresponding Y compounds below the Fermi level.

\subsection{Magnetization, Exchange Couplings and {\tc}}

\redtx{Magnetization in Pure compounds agrees well with experiment:}

\begin{table*}[tbp]
\caption{Calculated spin $M_s$, orbital $M_l$, and total $M_\text{t}$
  magnetization, exchanges $J_0$ , Curie temperature {\tc} estimated
  in the mean-field approximation, and magnetocrystalline anisotropy
  $K$ in various compounds. Unless specified, experimental
  magnetization and anisotropy $K$ values from previous studies were
  measured or evaluated for low temperature ($<5~$K). }
\label{tbl:mjk}
\bgroup
\def\arraystretch{1.3}
\begin{tabular}{lcccccccccccccccc}
\hline \hline
\mr{2}{*}{Compound} & $M_s$ & $M_l$ & $M_\text{t}$       & & \mc{5}{c}{$J_0$ (meV)}                   & & \tc          & \dtc  & & \mc{2}{c}{$K$}                        & \mr{2}{*}{Ref.} \\ \cline{2-4} \cline{6-10} \cline{12-13} \cline{15-16} 
   &       \mc{3}{c}{\mubfuB}                            & & 8$i$ (Ti)& 8$i$  &  8$j$ &  8$f$ & Y     & &    \mc{2}{c}{K}      & & \mevfuB       & \mjmcB                &  \\ \hline
{\yfeL}         &  24.20  &  0.61  &  24.81              & &          &  7.57 &  4.91 &  5.10 & 1.34  & & 689          & -38   & &  1.40         &  1.34                 &      \\
{\yfeti}~(Exp.) &         &        &  19$-$20.6          & &          & & & &                         & & 524$-$538    &       & &               &  2.0                  & [\onlinecite{yang.ssc1988,hu.jap1990,qi.jpcm1992,wang.jpcm2001,tereshina.jpcm2001}]  \\
{\yfeti}        &  19.75  &  0.60  &  20.35              & &  5.29    &  7.17 &  6.70 &  6.61 & 1.43  & & 727          &   0   & &  1.93         &  1.83                 &      \\
{\yfetih}       &  19.92  &  0.54  &  20.46              & &  4.99    &  7.63 &  7.46 &  7.57 & 1.40  & & 778          &  51   & &  2.07         &  1.96                 &      \\
{\yfetic}       &  20.64  &  0.55  &  21.19              & &  5.51    &  8.58 &  8.83 &  8.66 & 1.67  & & 884          & 157   & &  0.95         &  0.87                 &      \\
{\yfetin}       &  22.11  &  0.57  &  22.68              & &  5.44    &  9.36 &  8.91 &  9.29 & 1.30  & & 938          & 211   & &  1.80         &  1.65                 &      \\
{\ycoti}~(Exp.) &         &        &  14.2$-$15.7        & &          & & & &                         & & 1020$-$1050  &       & &               &  0.75\fk{1}           & [\onlinecite{wang.jpcm2001,ohashi.jap1991}]     \\
{\ycoti}        &  14.42  &  0.82  &  15.24              & &  3.93    & 10.50 & 10.13 & 11.13 & 1.45  & & 1091         & 364   & &  0.94         &  0.93                 &      \\
{\ycoL}         &  18.42  &  0.90  &  19.32              & &          & 12.52 & 13.12 & 13.84 & 1.66  & & 1374         & 647   & &  0.48         &  0.48                 &      \\
\hline                                                                                                                                                                   
{\cefeL}        &  24.02  &  0.78  &  24.80              & &          &  8.69 &  7.33 &  6.12 & 1.75  & & 806          & 131   & &  1.77         &  1.69                 &      \\
{\cefeti}~(Exp.)&         &        &  17.4$-$20.2        & &          & & & &                         & & 482$-$487    &       & &               &  1.3\fk{1}$-$2.0\fk{1}& [\onlinecite{ohashi.jap1991,pan.jap1994,isnard.jac1998,zhou.jap2014,goll.pss2014}]    \\
{\cefeti}       &  19.19  &  0.72  &  19.91              & &  4.69    &  6.26 &  7.04 &  5.95 & 2.16  & & 675          &   0   & &  2.09         &  1.98                 &      \\
{\cefetih}      &  20.24  &  0.77  &  21.01              & &  4.67    &  6.87 &  7.42 &  7.04 & 2.30  & & 736          &  61   & &  2.03         &  1.90                 &      \\
{\cefetic}      &  19.84  &  0.73  &  20.57              & &  5.45    &  9.86 &  8.62 &  8.62 & 3.44  & & 908          & 233   & &  1.09         &  0.99                 &      \\
{\cefetin}      &  21.48  &  0.67  &  22.15              & &  5.51    &  9.09 &  8.53 &  8.99 & 1.09  & & 905          & 230   & &  1.78         &  1.62                 &      \\
{\cecoti}~(Exp.)&         &        &  10.9$-$12.53\fk{1} & &          & & & &                         & & 920$-$937    &       & &  \mc{2}{c}{\text{Axial}}              & [\onlinecite{ohashi.jap1991,zhou.jap2014}] \\
{\cecoti}       &  13.77  &  1.32  &  15.09              & &  4.07    & 10.40 &  9.38 & 10.94 & 3.76  & & 1044         & 369   & &  1.29         &  1.23                 &      \\
{\cecoL}        &  17.35  &  1.36  &  18.71              & &          & 12.03 & 12.29 & 12.97 & 3.80  & & 1286         & 611   & &  1.24         &  1.24                 &      \\
\hline
\hline
\end{tabular}\\ 
\egroup
\footnotetext[1]{ Measured at room temperature.}
\end{table*}

Intrinsic magnetic properties of each compound are listed in
\rtbl{tbl:mjk}. Experimental magnetization and anisotropy values
vary. The calculated magnetizations in {\yfeti}, {\ycoti}, and
{\cefeti} compare well with experiments. For {\cecoti}, only a limited
number of studies had been reported, and the calculated magnetization
is larger than experimental ones.  Ti spin moments couple antiparallel
to those of Fe and Co sublattices, which is typical for the light $3d$
and $4d$ elements.\cite{zhao.prl2014} In {\cefeti}, the Ce spin
moments antiferromagnetically couple with the $TM$ sublattice as
expected.\cite{trygg.jmmm1992} Ce has a spin moment
{\mspin}$\approx$-0.7{\mub} and an orbital moment
{\morb}$\approx$0.3{\mub} with the opposite sign, which reflects
Hund's third rule. The calculated Fe spin moments on the individual
sublattice have the magnitude in the sequence of
{\mspin}($8i$)$>${\mspin}($8j$)$>${\mspin}($8f$), which agrees with
previous experiments and calculations.\cite{hu.jpcm1989} The dumbbell
8$i$ sites have larger spin magnetic moments because of the relative
larger surrounding empty volume and smaller atomic coordination
number. The orbital magnetic moments calculated are larger in the
Co-rich compounds than the Fe-rich compounds. MFA overestimated {\tc}
by about 200{\KK} in Fe compounds and about 50$-$100{\KK} in Co
compounds, respectively. RPA gives lower {\tc} values, e.g., 489{\KK}
in {\yfeti}, and 461{\KK} in {\cefeti}, respectively. The experimental
{\tc} falls between the MFA and RPA values, and is much closer to the
latter.

\redtx{Ti quickly decrease the magnetization but not {\tc}:}
Ti additions decrease the magnetization by $20\%$ in {\rfeti} and
{\rcoti} relative to their 1-12 hypothetical counterparts. The
magnetization reduction is not only due to the replacement of
ferromagnetic Fe by antiferromagnetic Ti atoms (spin moment
$-$0.54{\mub}), but also the suppression of the ferromagnetism on the
neighboring Fe sublattices. This is a common effect of doping early
3$d$ or 4$d$ elements on the Fe or Co sublattice.\cite{zhao.prl2014}
On the other hand, the addition of the Ti atom barely affects the Ce
moment. Interestingly, although magnetization decreased by $20\%$ upon
the Ti addition, the calculated {\tc} is even slightly higher in
{\yfeti} than in {\yfeL}. This is somewhat reflected in the
experiments, in which no obvious {\tc} dependence on Ti composition
was observed in YFe$_{11-z}$Ti$_{z}$ over the homogeneous 1-12 phase
composition range, 0.7$\leq z \leq$ 1.25.\cite{hu.jap1990}

To understand this phenomenon, we investigated the effective exchange
coupling parameters $J_{0}(i)$=$\sum_{j}'J_{ij}$ and compare $J_{0}$
values in {\yfeL} and {\yfeti}. With Ti replacing one Fe atom, $J_0$
values increase for all sites except the pair of Ti-Fe dumbbell
sites. The overall $J_{0}$ and the mean-field {\tc} increase. The
site-resolved effective exchange parameters $J_{0}(i)$ for various
compounds are listed in \rtbl{tbl:mjk}.

\redtx{ M vs Co agrees well with experiments:} 
\begin{figure}[tbp]
\begin{tabular}{cc}
\includegraphics[width=.45\textwidth,clip,angle=0]{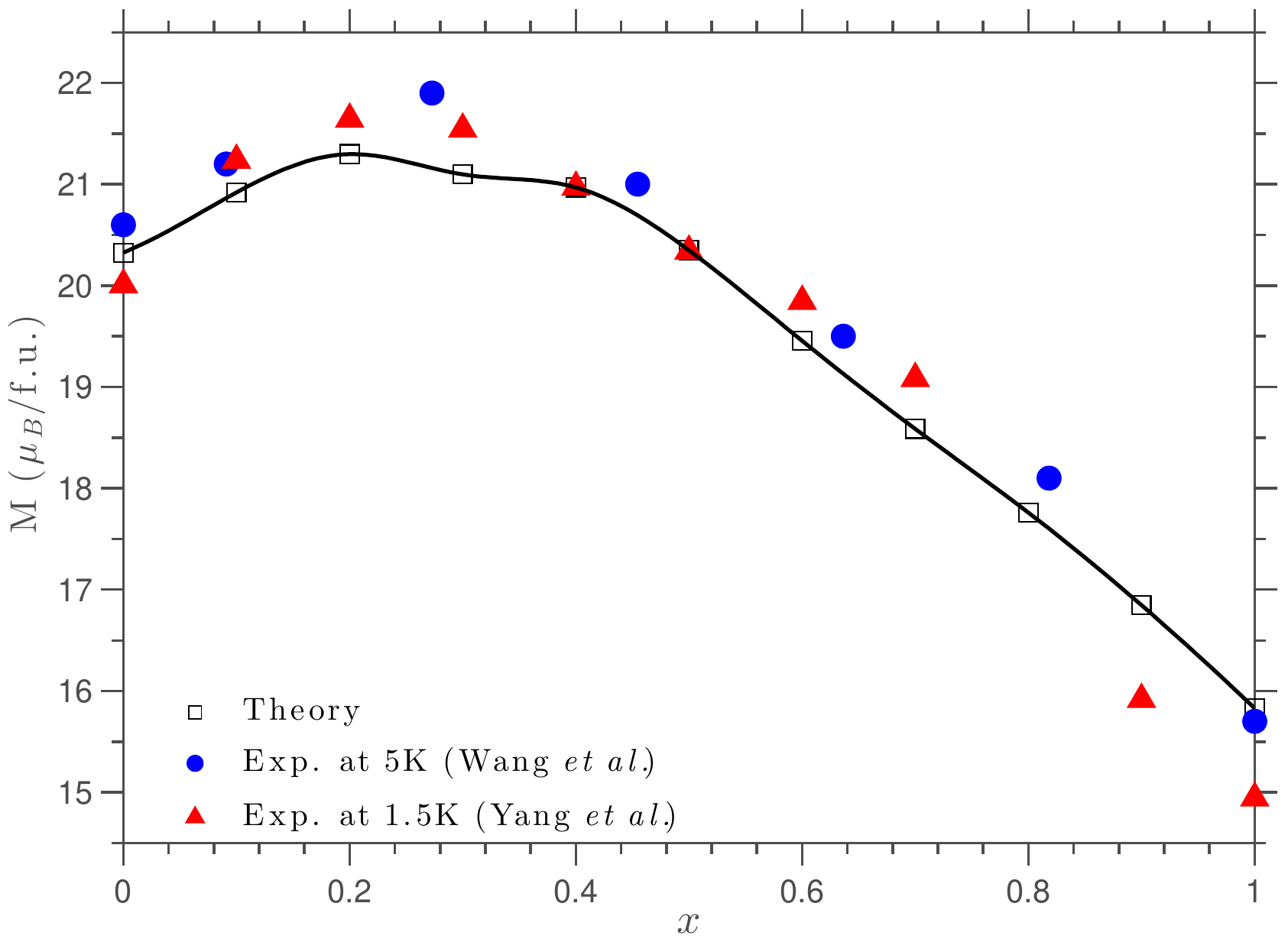} \\
\end{tabular}
\caption{ (Color online) Comparison of measured and calculated
  (squares) $M$ versus Co content in {\yfecoti}. Experimental data are
  from Wang~{\etal}\cite{wang.jpcm2001} at 5{\KK} (circles) and
  Yang~{\etal}\cite{yang.ssc1988} at 1.5{\KK} (triangles). }
\label{fig:mvsxco}
\end{figure}

Figure \ref{fig:mvsxco} shows the magnetization as a function of the
Co composition in {\yfeti}, with similar behavior to the
Slater-Pauling curve. The maximum magnetization occurs at $x$=0.2,
while in experiments it is at $x$=0.3.\cite{wang.jpcm2001} Similarly,
for {\cefecoti}, the experimental maximum magnetization occurs at
$x$=0.1$-$0.15.\cite{goll.pss2014} As shown in \rtbl{tbl:mjk}, the
{\rcoti} compounds have much larger {\tc} than the corresponding
{\rfeti} compounds, which agrees with experiments.\cite{zhou.jap2014}

\redtx{ H, C, and N doping, especially the later, increase M and {\tc}:}
All interstitial doping increases $M$ and {\tc} in {\yfeti} and
{\cefeti}, and nitriding has the strongest effect. With H, C, and N
doping, the calculated Curie temperature in {\yfeti} increases by 51,
157, and 211{\KK}, respectively, which is consistent with
experiments. $J_{0}$ values on all three $TM$ sublattices increase
with interstitial doping. Although DFT underestimats the volume
expansion with H doping, the calculated {\dtc} is only slightly
smaller than the experimental value.  The calculated {\dtc} is larger
with N doping than C doping, while their calculated volume expansions
are similar. This indicates that both volume and chemical effects are
important for the {\tc} enhancement. To estimate qualitatively the
relative magnitudes of the two effects, we calculate the {\tc} of
several hypothetical compounds related to {\yfeti}N by removing the N
atom in the unit cell or replacing it with H or C atoms,
respectively. The calculated {\dtc} of those structures relative to
{\yfeti} are 53, 80, and 169{\KK}, respectively. Obviously, both
volume and chemical effects contribute to the {\tc} enhancement and
the chemical effects of interstitial elements are in the sequence of
N$>$C$>$H.

\subsection{MAE in {\rfecoti}}

\redtx{MAE in pure compound agrees well with experiments:} 
As listed in \rtbl{tbl:mjk}, both {\yfeti} and {\ycoti} have uniaxial
anisotropy. Calculated MAE in {\yfeti} is in good agreement with the
experimental value. {\cefeti} has a slightly larger MAE than {\yfeti}
as found in experiments.\cite{coey.jap1993,isnard.jac1998} The PBE
functional (not shown) gives a smaller MAE than LDA in {\yfeti} and
{\cefeti}.

\redtx{What is interesting?}  
The Fe sublattice anisotropy may have a strong dependence on the
composition of stabilizer atoms.\cite{solzi.jap1988} To understand how
Ti affects the magnetic anisotropy and the origin of the nonmonotonic
dependence of MAE on Co composition, we resolved MAE into sites by
evaluating the matrix element of the on-site SOC
energy.\cite{ke.prb2015, antropov.ssc2014} For intermediate Co
composition, we investigate the MAE in YFe$_{7}$Co$_{4}$Ti and
YFe$_{3}$Co$_{8}$Ti. We calculated the formation energy relative to
{\yfeti} and {\ycoti} and found that YFe$_{7}$Co$_{4}$Ti has a
formation energy $E_\text{fmn}$=$-$34{\mevat} with four Co atoms on
the $8j$ sites and $E_\text{fmn}$=$-$28{\mevat} with four Co atoms on
the $8f$ sites. Both values are lower than
$E_\text{fmn}$=$-$10{\mevat}, the formation energy of YFe$_8$Co$_3$Ti
with all three Co atoms being on the $8i$ sites. Hence, the site
preference of Co atoms is 8$j$$>$8$f$$>$8$i$, which agrees with the
neutron scattering experiments.\cite{liang.jap1999} For
YFe$_{3}$Co$_{8}$Ti, we occupy another four Co atoms on the $8f$ sites
and the corresponding formation energy is $-$31{\mevat}.

\begin{figure}[tbp]
\begin{tabular}{cc}
\includegraphics[width=.45\textwidth,clip,angle=0]{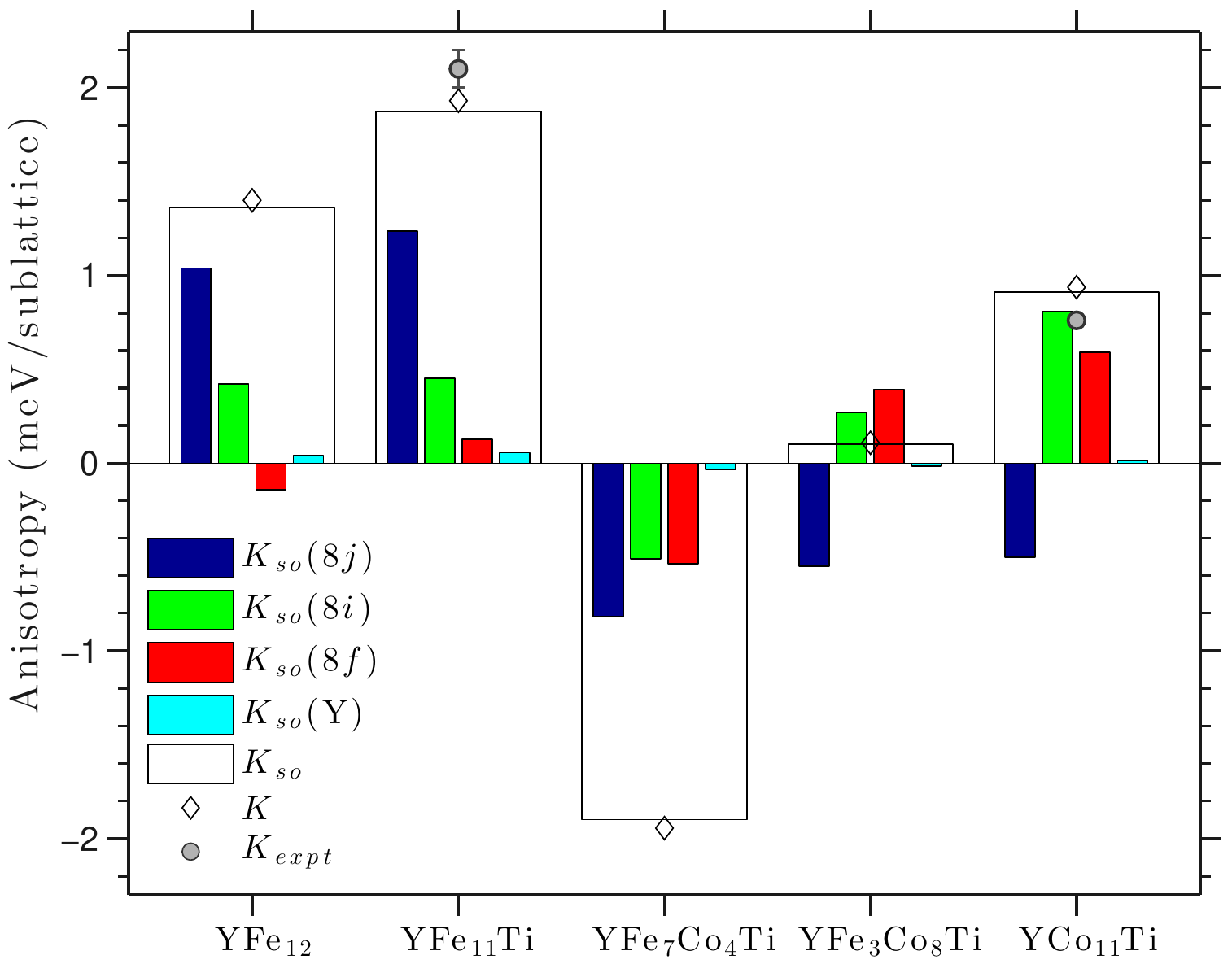} \\
\end{tabular}
\caption{ (Color online) Total and sublattice-resolved {\kso} in
  {\yfeti}, YFe$_{7}$Co$_{4}$Ti, YFe$_{3}$Co$_{8}$Ti and {\ycoti}.
  Calculated $K$ and measured $K_\text{expt}$ values are also
  compared. Experimental values were from
  Refs.\ \onlinecite{nikitin.ijhe1999} and
  \onlinecite{ohashi.jap1991}, measured at 4.2K for {\yfeti} and 293K
  for {\ycoti}, respectively.  In calculations, we assume that all
  four Co occupy the 8$j$ sites in YFe$_{7}$Co$_{4}$Ti while all eight
  Co occupy the 8$j$ and 8$f$ sites in YFe$_{3}$Co$_{8}$Ti.}
\label{fig:ksublattice}
\end{figure}

\redtx{Ti increase MAE:}

Figure~\ref{fig:ksublattice} shows the total MAE values and their
sublattice-resolved components, in YFe$_{12}$, YFe$_{7}$Co$_{4}$Ti,
YFe$_{3}$Co$_{8}$Ti and YCo$_{11}$Ti.  Obviously, \req{eq:eq1} is
well satisfied in all compounds and {\ksoc} presents well the
site-resolved MAE. The Y sublattice has a negligible contribution to
anisotropy, as expected for a weakly magnetic atom, because the
spin-parallel components of MAE contribution cancel out the spin-flip
ones.\cite{ke.prb2015} Sublattice-resolved MAE contributions in
{\yfeL} shows $K(8j)$$>$$K(8i)$$>$0$>$$K(8f)$, which agrees with the
previous estimation in sign but differs in the order.\cite{hu.jap1990}
Considering Fe(8$i$) sites have positive contributions to the uniaxial
anisotropy in {\yfeL}, one may expect that replacing Fe atoms by the
Ti atoms on the {$8i$} site would decrease MAE. Interestingly, we
found that {\yfeti} has even larger uniaxial anisotropy than
{\yfeL}. Anisotropies of all three sublattices become more uniaxial
and $K(8j)$$>$$K(8i)$$>$$K(8f)$$>$0 in {\yfeti}, which indicates that
the introduction of Ti atoms modifies the electronic structure of
neighboring sites and enhances their contribution to uniaxial
anisotropy. Similarly, other compounds, such as {\ycoti}, {\cefeti},
and {\cecoti}, are also found to have MAE values larger than or
similar to their corresponding hypothetical 1-12 counterparts.

\redtx{Non monotonic MAE behavior can be understood with site preference:}
The dependence of MAE on the Co composition is nonmonotonic and also
found in other $R$-$TM$ systems.\cite{thuy.jpc1988} As shown in
\rfig{fig:ksublattice}, the calculated MAE reproduce the trend
observed in experiment. For intermediate Co compositions,
YFe$_{7}$Co$_{4}$Ti compound has planar anisotropy while
YFe$_{3}$Co$_{8}$Ti compound has a very small uniaxial anisotropy. The
8$j$ sublattice is the major contributor to the uniaxial anisotropy in
{\yfeti}. With all four 8$j$ Fe atoms being replaced by Co atoms in
YFe$_{7}$Co$_{4}$Ti, $K(8j)$ becomes very negative. Moreover, $K(8i)$
and $K(8f)$ are also strongly affected and become negative. Further Co
doping on $8f$ sites changes $K(8i)$ and $K(8f)$ back to positive in
YFe$_{3}$Co$_{8}$Ti. Finally, in {\ycoti} both $K(8i)$ and $K(8f)$
increase and $K(8j)$ becomes less planar and we have
$K(8i)$$>$$K(8f)$$>$0$>$$K(8j)$.

The nonmonotonic composition dependence is often interpreted by
preferential site occupancy,\cite{thuy.jpc1988} however, such an
explanation is an oversimplification for a metallic system, such as
{\yfecoti}. The MAE contributions from each $TM$ sublattice may depend
on the detailed band structure around the Fermi energy. The doping of
Co on particular sites unavoidably affects the electronic structure of
neighboring $TM$ sublattices due to the hybridization between them,
which changes the MAE contribution from neighboring sites. Obviously,
as shown in \rfig{fig:ksublattice}, with a sizable amount of Co
doping, the variation of anisotropy is a collective effect instead of
a sole contribution from the doping sites.

Among three $TM$ sublattices, the dumbbell 8$i$ sites have the largest
contribution to the uniaxial anisotropy in {\ycoti}, which we found
also true in {\cecoti}, and hypothetical {\ycoL} and {\cecoL}. It is
interesting to compare the MAE contributions from Co sublattices in
$R$Co$_{12}$ and $R_2$Co$_{17}$, in which the dumbbell Co sites have
the most negative contribution to the uniaxial
anisotropy.\cite{ke.intermag2015} In both cases, the moments of the
dumbbell sites prefer to be perpendicular to the dumbbell bonds, which
are along different directions in two structures, i.e., basal axes in
the 1-12 structure and $c$ axis in the 2-17 structure. As a result,
dumbbell Co sites have MAE contributions of opposite sign in two
structures.

\begin{figure}[tbp]
\begin{tabular}{cc}
\includegraphics[width=.45\textwidth,clip,angle=0]{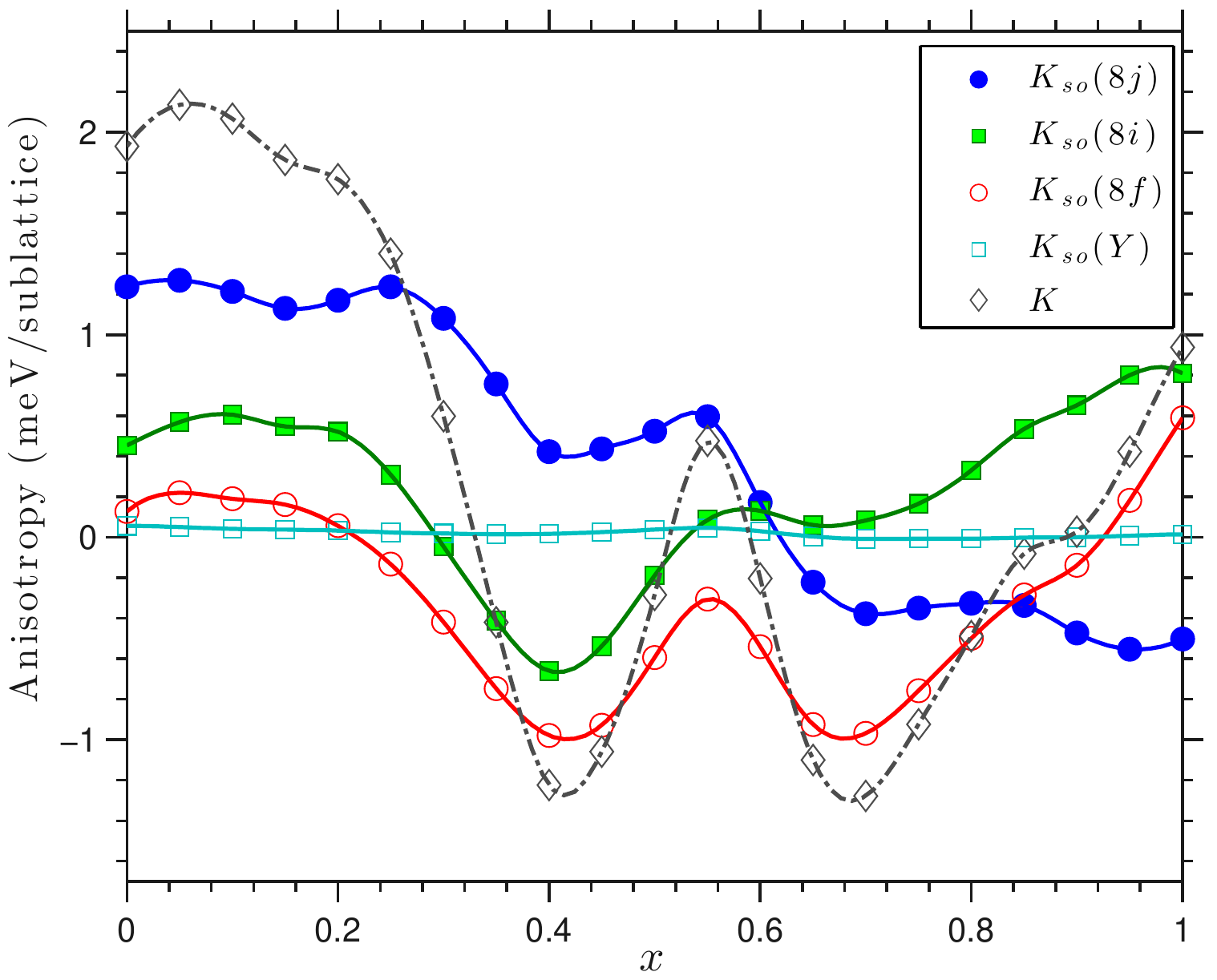} \\
\includegraphics[width=.45\textwidth,clip,angle=0]{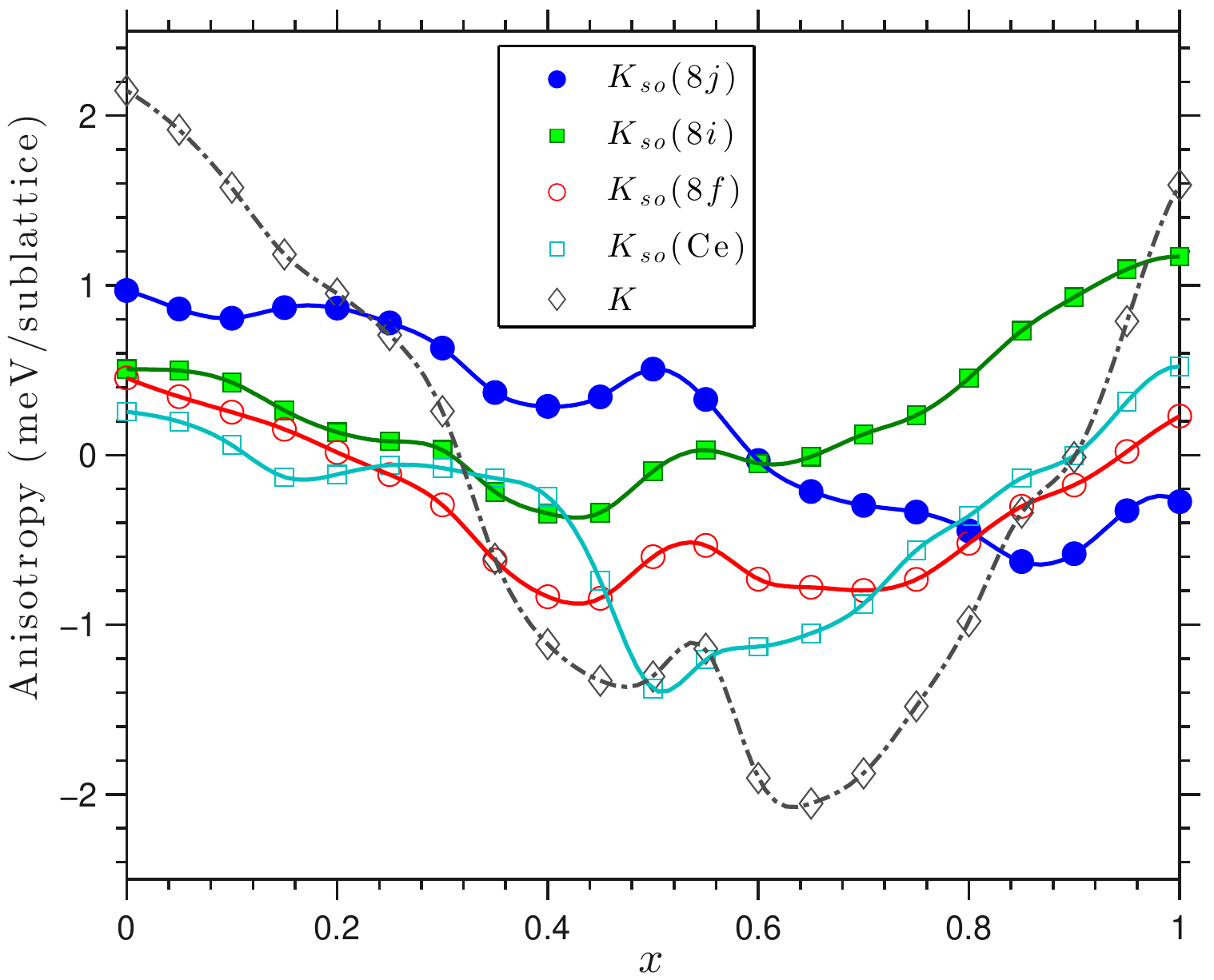} \\
\end{tabular}
\caption{ (Color online) $K$ and sublattice-resolved {\ksoc} in
  Y-based (top) and Ce-based (bottom)
  $R$(Fe$_{1-x}$Co$_{x}$)$_{11}$Ti.}
\label{fig:ksosite_vca}
\end{figure}

In a real sample, Co likely also partially occupies the $8j$ and $8f$
sites instead of exclusively only the 8$j$ site. We investigate the
scenario at the other extreme by assuming Co occupies the three $TM$
sublattices with equal probability and calculate composition
dependence of MAE using the virtual crystal approximation (VCA).
Interestingly, the nonmonotonic behavior is also observed as shown in
\rfig{fig:ksosite_vca}. The easy direction changes from uniaxial to
in-plane and then back to uniaxial. The variations of each individual
$TM$ sublattice share a similarity with the trend shown in
\rfig{fig:ksublattice}. With increasing of $x$ in {\yfecoti}, $K(8j)$
decreases and becomes negative while $K(8i)$ and $K(8f)$ become
negative for the intermediate Co composition and then change back to
positive at the Co-rich end. Thus, the nonmonotonic behavior is
confirmed with or without considering preferential occupancy. The
spin-reorientation transition\cite{cheng.jap1991} from axis to
in-plane occurs in {\yfecoti} but not pure
{\yfeti},\cite{cheng.jap1991} which may relate to the fact that the
competing anisotropies between three $TM$ sublattices exist in
{\yfecoti} while all three $TM$ sublattices support the uniaxial
anisotropy in {\yfeti}. As shown in \rfig{fig:ksosite_vca}(Top), MAE
in {\yfecoti} barely changes or even slightly increases with a very
small Co composition. A similar feature had been observed
experimentally.\cite{tereshina.jac2005} It is caused by the partial
occupation of Co on $8f$ sites in {\yfeti}. We found that replacing Fe
atoms in {\yfeti} with Co atoms on the 8$f$ sites increases the MAE.

\redtx{Ce compound:}
It is commonly assumed that the MAE contributions from the $TM$
sublattices are similar in $R$-$TM$ compounds with different $R$, and
such contributions are often estimated experimentally from
measurements on corresponding yttrium compounds.\cite{coey.jmmm1989}
As shown in \rfig{fig:ksosite_vca}, MAE contributions from $TM$
sublattices in {\yfeti} and {\cefeti} are similar but not
identical. All three $TM$ sublattices have positive contributions to
the uniaxial anisotropy and $K(8j)$$>$$K(8i)$$>$$K(8f)$$>$0. However,
magnitudes of each sublattice differ in two compounds, which suggest
that the hybridization $TM$ sites have with different $R$ atoms
affects their contributions to the MAE. Unlike the Y sublattice in
{\yfeti}, Ce provides a positive contribution to the uniaxial
anisotropy in {\cefeti}.

\subsection{Effect of interstitial doping}
\begin{figure}[tbp]
\begin{tabular}{cc}
\includegraphics[width=0.45\textwidth,clip,angle=0]{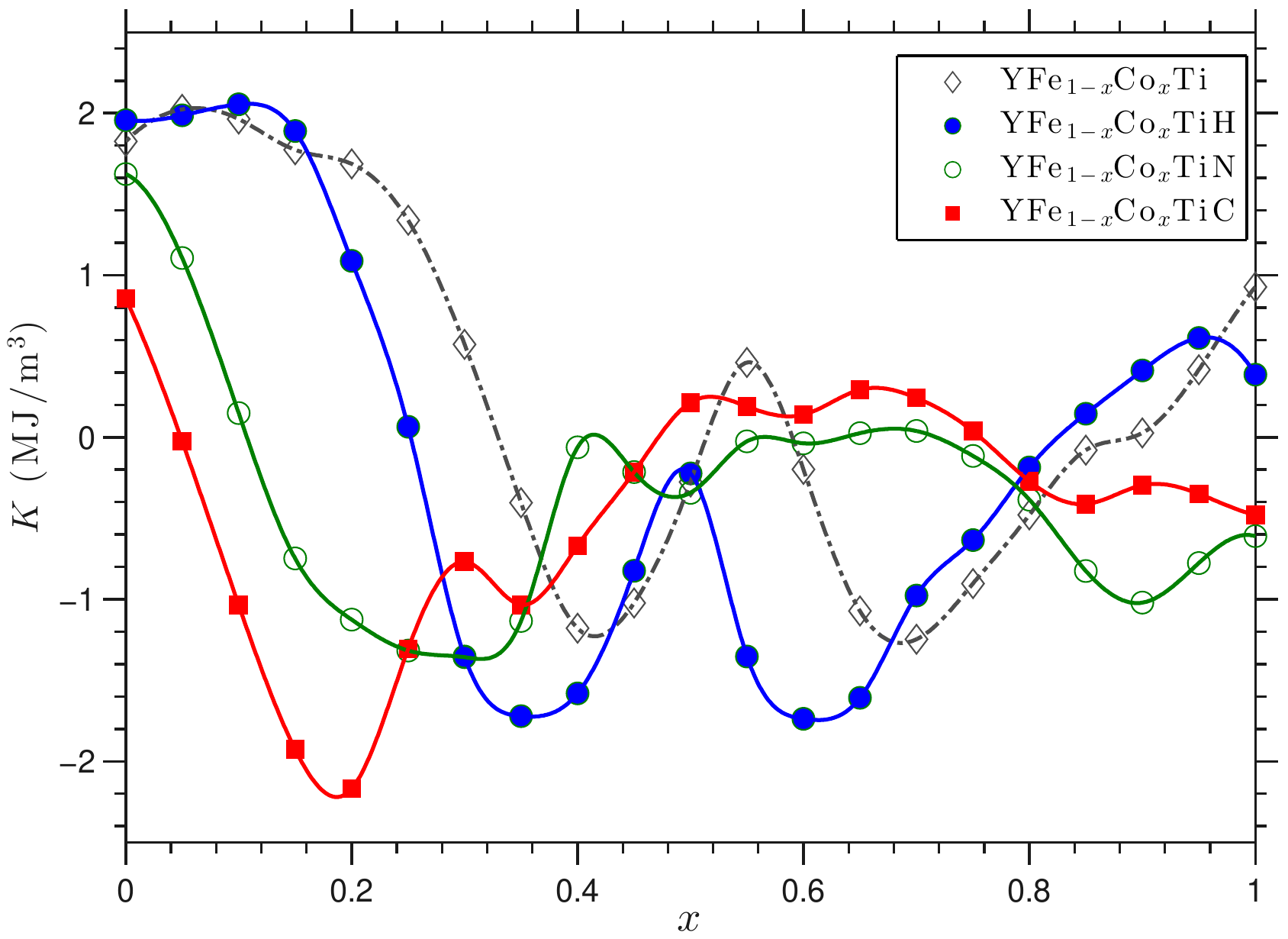} \\
\includegraphics[width=0.45\textwidth,clip,angle=0]{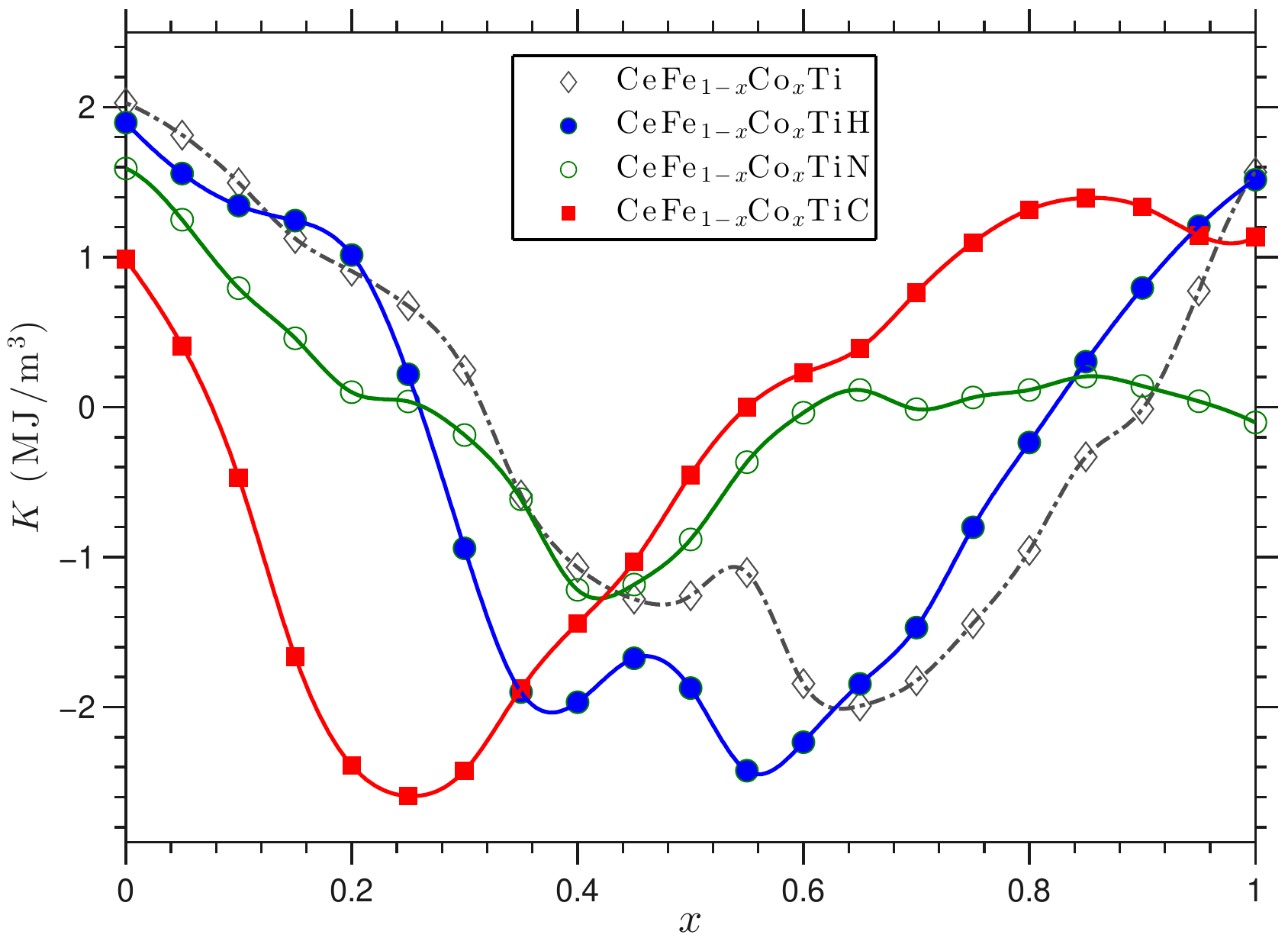} \\
\end{tabular}
\caption{ (Color online) $K$ versus Co content in
  $R$(Fe$_{1-x}$Co$_{x}$)$_{11}$Ti$Z$ with $R$=Y (top) and $R$=Ce
  (bottom), with and without $Z$=H, C, and N. }
\label{fig:khbcn}
\end{figure}

\redtx{H, C, and N doping on {\rfecoti}:} 
Interstitial doping with N, C, and H affects the MAE from both of the
Fe and $R$ sublattices.\cite{nikitin.jac2001} As shown in
\rtbl{tbl:mjk}, H doping barely changes or slightly increases the
uniaxial anisotropy in {\yfeti} and {\cefeti} while carbonizing and
nitriding weaken the uniaxial anisotropy, which agrees with
experiments.\cite{qi.jpcm1992,pan.jap1994} Simultaneous substitutional
Co doping and interstitial doping with H, C, or N is of interest.
Although the uniaxial anisotropy may not improve that much at the low
temperature, the effect could be more significant at room
temperature. For example, upon hydrogenation, a significant increase
of $K_1$ with a factor 1.8 was observed in YFe$_9$Co$_2$Ti at room
temperature.\cite{tereshina.jac2005}

To our knowledge, simultaneous doping of Co and interstitial elements
C and N atoms is not well studied. We calculated the MAE dependence on
Co compositions in {\cefecotiz} with $Z$=H, C, and N, and results are
shown in \rfig{fig:khbcn}. The site preference of Co is not considered
and VCA is used. The maximum of uniaxial anisotropy in {\yfecoti}H is
obtained at $x$=0.1 while experiments found the maximum at
YFe$_9$Co$_2$TiH.\cite{tereshina.jac2005} For the Fe-rich
{\cecoti}$Z$, only H doping slightly increases the MAE, while C and N
quickly decrease uniaxial anisotropy. For {\yfecotiz}, it is unlikely
we can have better uniaxial anisotropy (at least at low temperature)
over the whole range of Co composition.  Interestingly, for Co-rich
{\cefecotiz}, interstitial C doping significantly improves the
uniaxial anisotropy in {\cefecotiz} for 0.7$<$$x$$<$0.9. Considering
the relative high Curie temperature on the Co-rich end, it has an
attractive combination of all three intrinsic magnetic properties,
$M$, $J$, and $K$, for permanent magnet application.
 
\section{Conclusion}
\label{sec:4}
Using DFT methods, the intrinsic magnetic properties of
{\rfeti}-related systems were investigated for the effects of
substitutional alloying with Co and interstitial doping with H, C, and
N. All properties and trends were well described within the local
density approximation to DFT.  In comparison to the hypothetical
{\yfeL}, Ti quickly decreases the magnetization and increases the
uniaxial magnetic anisotropy in {\yfeti}.  The calculated Co site
preference is $8j>8f>8i$ in {\yfecoti} with $x<0.4$, in agreement with
neutron experiments.  The enhancement of $M$ and {\tc} due to Co
doping and interstitial doping are in good agreement with experiments.

Compared with {\yfeti}, the calculated {\tc} increases by 51, 157 and
211{\KK} in {\yfeti}$Z$ with $Z$=H, C, and N, respectively, with both
volume and chemical effects contributing to the enhancement.  We found
that all three Fe sublattices promote uniaxial anisotropy in the
sequence of ${K(8j)}>{K(8i)}>{K(8f)}>0$ in {\yfeti}, while competing
contributions give ${K(8i)}>{K(8f)}>{0}>{K(8j)}$ in {\ycoti}.  For
intermediate Co composition, we confirm that the easy direction
changes with increasing Co content from uniaxial to in-plane and then
back to uniaxial. Substitutional doping affects the MAE contributions
from neighboring sites and the nonmonotonic composition dependence of
anisotropy is a collective effect, which can not be solely explained
by preferential occupancy. The Ce sublattice promotes the uniaxial
anisotropy in {\cefeti} and {\cecoti}. Interstitial C doping
significantly increases the uniaxial anisotropy in {\cefecoti} for
$0.7<x<0.9$, which may provide the best combination of all three
intrinsic magnetic properties for permanent applications.

\section*{Acknowledgments}
We thank B. Harmon, A. Alam, C. Zhou, and R. W. McCallum for helpful
discussions.  This work was supported by the U.S. Department of Energy
ARPA-E (REACT 0472-1526).  Ames Laboratory is operated for the
U.S. DOE by Iowa State University under Contract
No. DE-AC02-07CH11358.

\bibliography{aaa}

\begin{thebibliography}{60}
\expandafter\ifx\csname natexlab\endcsname\relax\def\natexlab#1{#1}\fi
\expandafter\ifx\csname bibnamefont\endcsname\relax
  \def\bibnamefont#1{#1}\fi
\expandafter\ifx\csname bibfnamefont\endcsname\relax
  \def\bibfnamefont#1{#1}\fi
\expandafter\ifx\csname citenamefont\endcsname\relax
  \def\citenamefont#1{#1}\fi
\expandafter\ifx\csname url\endcsname\relax
  \def\url#1{\texttt{#1}}\fi
\expandafter\ifx\csname urlprefix\endcsname\relax\def\urlprefix{URL }\fi
\providecommand{\bibinfo}[2]{#2}
\providecommand{\eprint}[2][]{\url{#2}}

\bibitem[{\citenamefont{McCallum et~al.}(2014)\citenamefont{McCallum, Lewis,
  Skomski, Kramer, and Anderson}}]{mccallum.arms2014}
\bibinfo{author}{\bibfnamefont{R.}~\bibnamefont{McCallum}},
  \bibinfo{author}{\bibfnamefont{L.}~\bibnamefont{Lewis}},
  \bibinfo{author}{\bibfnamefont{R.}~\bibnamefont{Skomski}},
  \bibinfo{author}{\bibfnamefont{M.}~\bibnamefont{Kramer}}, \bibnamefont{and}
  \bibinfo{author}{\bibfnamefont{I.}~\bibnamefont{Anderson}},
  \bibinfo{journal}{Annual Review of Materials Research}
  \textbf{\bibinfo{volume}{44}}, \bibinfo{pages}{451} (\bibinfo{year}{2014}).

\bibitem[{\citenamefont{Kusne et~al.}(2014)\citenamefont{Kusne, Gao, Mehta, Ke,
  Nguyen, Ho, Antropov, Wang, Kramer, Long et~al.}}]{kusne.srep2014}
\bibinfo{author}{\bibfnamefont{A.}~\bibnamefont{Kusne}},
  \bibinfo{author}{\bibfnamefont{T.}~\bibnamefont{Gao}},
  \bibinfo{author}{\bibfnamefont{A.}~\bibnamefont{Mehta}},
  \bibinfo{author}{\bibfnamefont{L.}~\bibnamefont{Ke}},
  \bibinfo{author}{\bibfnamefont{M.}~\bibnamefont{Nguyen}},
  \bibinfo{author}{\bibfnamefont{K.}~\bibnamefont{Ho}},
  \bibinfo{author}{\bibfnamefont{V.}~\bibnamefont{Antropov}},
  \bibinfo{author}{\bibfnamefont{C.}~\bibnamefont{Wang}},
  \bibinfo{author}{\bibfnamefont{M.}~\bibnamefont{Kramer}},
  \bibinfo{author}{\bibfnamefont{C.}~\bibnamefont{Long}}, \bibnamefont{et~al.},
  \bibinfo{journal}{{Sci. Rep.}} \textbf{\bibinfo{volume}{4}}
  (\bibinfo{year}{2014}).

\bibitem[{\citenamefont{Yang et~al.}(1988{\natexlab{a}})\citenamefont{Yang,
  Kong, Sun, Gu, and Cheng}}]{yang.jap1988}
\bibinfo{author}{\bibfnamefont{Y.}~\bibnamefont{Yang}},
  \bibinfo{author}{\bibfnamefont{L.}~\bibnamefont{Kong}},
  \bibinfo{author}{\bibfnamefont{S.}~\bibnamefont{Sun}},
  \bibinfo{author}{\bibfnamefont{D.}~\bibnamefont{Gu}}, \bibnamefont{and}
  \bibinfo{author}{\bibfnamefont{B.}~\bibnamefont{Cheng}},
  \bibinfo{journal}{Journal of Applied Physics} \textbf{\bibinfo{volume}{63}},
  \bibinfo{pages}{3702} (\bibinfo{year}{1988}{\natexlab{a}}).

\bibitem[{\citenamefont{Buschow}(1991)}]{buschow.jmmm1991}
\bibinfo{author}{\bibfnamefont{K.}~\bibnamefont{Buschow}},
  \bibinfo{journal}{Journal of Magnetism and Magnetic Materials}
  \textbf{\bibinfo{volume}{100}}, \bibinfo{pages}{79 } (\bibinfo{year}{1991}).

\bibitem[{\citenamefont{Pan et~al.}(1994)\citenamefont{Pan, Liu, and
  Yang}}]{pan.jap1994}
\bibinfo{author}{\bibfnamefont{Q.}~\bibnamefont{Pan}},
  \bibinfo{author}{\bibfnamefont{Z.}~\bibnamefont{Liu}}, \bibnamefont{and}
  \bibinfo{author}{\bibfnamefont{Y.}~\bibnamefont{Yang}},
  \bibinfo{journal}{Journal of Applied Physics} \textbf{\bibinfo{volume}{76}},
  \bibinfo{pages}{6728} (\bibinfo{year}{1994}).

\bibitem[{\citenamefont{Sakuma}(1992)}]{sakuma.jpsj1992}
\bibinfo{author}{\bibfnamefont{A.}~\bibnamefont{Sakuma}},
  \bibinfo{journal}{Journal of the Physical Society of Japan}
  \textbf{\bibinfo{volume}{61}}, \bibinfo{pages}{4119} (\bibinfo{year}{1992}).

\bibitem[{\citenamefont{K{\"o}rner et~al.}(2016)\citenamefont{K{\"o}rner,
  Krugel, and Els{\"a}sser}}]{korner.sr2016}
\bibinfo{author}{\bibfnamefont{W.}~\bibnamefont{K{\"o}rner}},
  \bibinfo{author}{\bibfnamefont{G.}~\bibnamefont{Krugel}}, \bibnamefont{and}
  \bibinfo{author}{\bibfnamefont{C.}~\bibnamefont{Els{\"a}sser}},
  \bibinfo{journal}{Sci Rep} \textbf{\bibinfo{volume}{6}},
  \bibinfo{pages}{24686} (\bibinfo{year}{2016}).

\bibitem[{\citenamefont{Yang et~al.}(1988{\natexlab{b}})\citenamefont{Yang,
  Sun, Zhang, Luo, and Gao}}]{yang.ssc1988}
\bibinfo{author}{\bibfnamefont{Y.}~\bibnamefont{Yang}},
  \bibinfo{author}{\bibfnamefont{H.}~\bibnamefont{Sun}},
  \bibinfo{author}{\bibfnamefont{Z.}~\bibnamefont{Zhang}},
  \bibinfo{author}{\bibfnamefont{T.}~\bibnamefont{Luo}}, \bibnamefont{and}
  \bibinfo{author}{\bibfnamefont{J.}~\bibnamefont{Gao}},
  \bibinfo{journal}{Solid State Communications} \textbf{\bibinfo{volume}{68}},
  \bibinfo{pages}{175 } (\bibinfo{year}{1988}{\natexlab{b}}).

\bibitem[{\citenamefont{Franse and Radwa\'{n}ski}(1993)}]{hmm.v07c5}
\bibinfo{author}{\bibfnamefont{J.}~\bibnamefont{Franse}} \bibnamefont{and}
  \bibinfo{author}{\bibfnamefont{R.}~\bibnamefont{Radwa\'{n}ski}}
  (\bibinfo{publisher}{Elsevier, Amsterdam}, \bibinfo{year}{1993}),
  vol.~\bibinfo{volume}{7} of \emph{\bibinfo{series}{Handbook of Magnetic
  Materials}}, pp. \bibinfo{pages}{307 -- 501}.

\bibitem[{\citenamefont{Li and Coey}(1991)}]{hmm.v06c1}
\bibinfo{author}{\bibfnamefont{H.}~\bibnamefont{Li}} \bibnamefont{and}
  \bibinfo{author}{\bibfnamefont{J.}~\bibnamefont{Coey}}
  (\bibinfo{publisher}{Elsevier, Amsterdam}, \bibinfo{year}{1991}),
  vol.~\bibinfo{volume}{6} of \emph{\bibinfo{series}{Handbook of Magnetic
  Materials}}, pp. \bibinfo{pages}{1 -- 83}.

\bibitem[{\citenamefont{Hu et~al.}(1990)\citenamefont{Hu, Li, and
  Coey}}]{hu.jap1990}
\bibinfo{author}{\bibfnamefont{B.}~\bibnamefont{Hu}},
  \bibinfo{author}{\bibfnamefont{H.}~\bibnamefont{Li}}, \bibnamefont{and}
  \bibinfo{author}{\bibfnamefont{J.}~\bibnamefont{Coey}},
  \bibinfo{journal}{Journal of Applied Physics} \textbf{\bibinfo{volume}{67}},
  \bibinfo{pages}{4838} (\bibinfo{year}{1990}).

\bibitem[{\citenamefont{Coey}(1989)}]{coey.jmmm1989}
\bibinfo{author}{\bibfnamefont{J.}~\bibnamefont{Coey}},
  \bibinfo{journal}{Journal of Magnetism and Magnetic Materials}
  \textbf{\bibinfo{volume}{80}}, \bibinfo{pages}{9 } (\bibinfo{year}{1989}).

\bibitem[{\citenamefont{De~Boer et~al.}(1987)\citenamefont{De~Boer, Huang,
  De~Mooij, and Buschow}}]{deboer.jlcm1987}
\bibinfo{author}{\bibfnamefont{F.}~\bibnamefont{De~Boer}},
  \bibinfo{author}{\bibfnamefont{Y.}~\bibnamefont{Huang}},
  \bibinfo{author}{\bibfnamefont{D.}~\bibnamefont{De~Mooij}}, \bibnamefont{and}
  \bibinfo{author}{\bibfnamefont{K.}~\bibnamefont{Buschow}},
  \bibinfo{journal}{Journal of the Less Common Metals}
  \textbf{\bibinfo{volume}{135}}, \bibinfo{pages}{199 } (\bibinfo{year}{1987}).

\bibitem[{\citenamefont{Solzi et~al.}(1988)\citenamefont{Solzi, Pareti, Moze,
  and David}}]{solzi.jap1988}
\bibinfo{author}{\bibfnamefont{M.}~\bibnamefont{Solzi}},
  \bibinfo{author}{\bibfnamefont{L.}~\bibnamefont{Pareti}},
  \bibinfo{author}{\bibfnamefont{O.}~\bibnamefont{Moze}}, \bibnamefont{and}
  \bibinfo{author}{\bibfnamefont{W.}~\bibnamefont{David}},
  \bibinfo{journal}{Journal of applied physics} \textbf{\bibinfo{volume}{64}},
  \bibinfo{pages}{5084} (\bibinfo{year}{1988}).

\bibitem[{\citenamefont{Isnard et~al.}(1998)\citenamefont{Isnard, Miraglia,
  Guillot, and Fruchart}}]{isnard.jac1998}
\bibinfo{author}{\bibfnamefont{O.}~\bibnamefont{Isnard}},
  \bibinfo{author}{\bibfnamefont{S.}~\bibnamefont{Miraglia}},
  \bibinfo{author}{\bibfnamefont{M.}~\bibnamefont{Guillot}}, \bibnamefont{and}
  \bibinfo{author}{\bibfnamefont{D.}~\bibnamefont{Fruchart}},
  \bibinfo{journal}{Journal of Alloys and Compounds}
  \textbf{\bibinfo{volume}{275-277}}, \bibinfo{pages}{637 }
  (\bibinfo{year}{1998}).

\bibitem[{\citenamefont{Tereshina et~al.}(2001)\citenamefont{Tereshina,
  Gaczyński, Rusakov, Drulis, Nikitin, Suski, Tristan, and
  Palewski}}]{tereshina.jpcm2001}
\bibinfo{author}{\bibfnamefont{I.}~\bibnamefont{Tereshina}},
  \bibinfo{author}{\bibfnamefont{P.}~\bibnamefont{Gaczyński}},
  \bibinfo{author}{\bibfnamefont{V.}~\bibnamefont{Rusakov}},
  \bibinfo{author}{\bibfnamefont{H.}~\bibnamefont{Drulis}},
  \bibinfo{author}{\bibfnamefont{S.}~\bibnamefont{Nikitin}},
  \bibinfo{author}{\bibfnamefont{W.}~\bibnamefont{Suski}},
  \bibinfo{author}{\bibfnamefont{N.}~\bibnamefont{Tristan}}, \bibnamefont{and}
  \bibinfo{author}{\bibfnamefont{T.}~\bibnamefont{Palewski}},
  \bibinfo{journal}{Journal of Physics: Condensed Matter}
  \textbf{\bibinfo{volume}{13}}, \bibinfo{pages}{8161} (\bibinfo{year}{2001}).

\bibitem[{\citenamefont{Zhou et~al.}(2014)\citenamefont{Zhou, Pinkerton, and
  Herbst}}]{zhou.jap2014}
\bibinfo{author}{\bibfnamefont{C.}~\bibnamefont{Zhou}},
  \bibinfo{author}{\bibfnamefont{F.}~\bibnamefont{Pinkerton}},
  \bibnamefont{and} \bibinfo{author}{\bibfnamefont{J.}~\bibnamefont{Herbst}},
  \bibinfo{journal}{Journal of Applied Physics} \textbf{\bibinfo{volume}{115}},
  \bibinfo{pages}{17C716} (\bibinfo{year}{2014}).

\bibitem[{\citenamefont{Ohashi et~al.}(1991)\citenamefont{Ohashi, Ido, Konno,
  and Yoneda}}]{ohashi.jap1991}
\bibinfo{author}{\bibfnamefont{K.}~\bibnamefont{Ohashi}},
  \bibinfo{author}{\bibfnamefont{H.}~\bibnamefont{Ido}},
  \bibinfo{author}{\bibfnamefont{K.}~\bibnamefont{Konno}}, \bibnamefont{and}
  \bibinfo{author}{\bibfnamefont{Y.}~\bibnamefont{Yoneda}},
  \bibinfo{journal}{Journal of applied physics} \textbf{\bibinfo{volume}{70}},
  \bibinfo{pages}{5986} (\bibinfo{year}{1991}).

\bibitem[{\citenamefont{Wang et~al.}(2001{\natexlab{a}})\citenamefont{Wang,
  Tang, Fuquan, Wang, Wang, Wu, and Yang}}]{wang.jpcm2001}
\bibinfo{author}{\bibfnamefont{J.}~\bibnamefont{Wang}},
  \bibinfo{author}{\bibfnamefont{N.}~\bibnamefont{Tang}},
  \bibinfo{author}{\bibfnamefont{B.}~\bibnamefont{Fuquan}},
  \bibinfo{author}{\bibfnamefont{W.}~\bibnamefont{Wang}},
  \bibinfo{author}{\bibfnamefont{W.}~\bibnamefont{Wang}},
  \bibinfo{author}{\bibfnamefont{G.}~\bibnamefont{Wu}}, \bibnamefont{and}
  \bibinfo{author}{\bibfnamefont{F.}~\bibnamefont{Yang}},
  \bibinfo{journal}{Journal of Physics: Condensed Matter}
  \textbf{\bibinfo{volume}{13}}, \bibinfo{pages}{1617}
  (\bibinfo{year}{2001}{\natexlab{a}}).

\bibitem[{\citenamefont{Sinha et~al.}(1989)\citenamefont{Sinha, Cheng, Wallace,
  and Sankar}}]{sinha.jmmm1989}
\bibinfo{author}{\bibfnamefont{V.}~\bibnamefont{Sinha}},
  \bibinfo{author}{\bibfnamefont{S.}~\bibnamefont{Cheng}},
  \bibinfo{author}{\bibfnamefont{W.}~\bibnamefont{Wallace}}, \bibnamefont{and}
  \bibinfo{author}{\bibfnamefont{S.}~\bibnamefont{Sankar}},
  \bibinfo{journal}{Journal of magnetism and magnetic materials}
  \textbf{\bibinfo{volume}{81}}, \bibinfo{pages}{227} (\bibinfo{year}{1989}).

\bibitem[{\citenamefont{Cheng et~al.}(1991)\citenamefont{Cheng, Sinha, Ma,
  Sankar, and Wallace}}]{cheng.jap1991}
\bibinfo{author}{\bibfnamefont{S.}~\bibnamefont{Cheng}},
  \bibinfo{author}{\bibfnamefont{V.}~\bibnamefont{Sinha}},
  \bibinfo{author}{\bibfnamefont{B.}~\bibnamefont{Ma}},
  \bibinfo{author}{\bibfnamefont{S.}~\bibnamefont{Sankar}}, \bibnamefont{and}
  \bibinfo{author}{\bibfnamefont{W.}~\bibnamefont{Wallace}},
  \bibinfo{journal}{Journal of Applied Physics} \textbf{\bibinfo{volume}{69}},
  \bibinfo{pages}{5605} (\bibinfo{year}{1991}).

\bibitem[{\citenamefont{Moze et~al.}(1995)\citenamefont{Moze, Pareti, and
  Buschow}}]{moze.jpcm1995}
\bibinfo{author}{\bibfnamefont{O.}~\bibnamefont{Moze}},
  \bibinfo{author}{\bibfnamefont{L.}~\bibnamefont{Pareti}}, \bibnamefont{and}
  \bibinfo{author}{\bibfnamefont{K.}~\bibnamefont{Buschow}},
  \bibinfo{journal}{Journal of Physics: Condensed Matter}
  \textbf{\bibinfo{volume}{7}}, \bibinfo{pages}{9255} (\bibinfo{year}{1995}).

\bibitem[{\citenamefont{Wang et~al.}(1999)\citenamefont{Wang, Fuquan, Yang, and
  Yang}}]{wang.jmsj1999}
\bibinfo{author}{\bibfnamefont{J.}~\bibnamefont{Wang}},
  \bibinfo{author}{\bibfnamefont{B.}~\bibnamefont{Fuquan}},
  \bibinfo{author}{\bibfnamefont{C.}~\bibnamefont{Yang}}, \bibnamefont{and}
  \bibinfo{author}{\bibfnamefont{F.}~\bibnamefont{Yang}},
  \bibinfo{journal}{Journal of the Magnetics Society of Japan}
  \textbf{\bibinfo{volume}{23}}, \bibinfo{pages}{459} (\bibinfo{year}{1999}).

\bibitem[{\citenamefont{Yang et~al.}(1991{\natexlab{a}})\citenamefont{Yang,
  Zhang, Kong, Pan, and Ge}}]{yang.apl1991}
\bibinfo{author}{\bibfnamefont{Y.}~\bibnamefont{Yang}},
  \bibinfo{author}{\bibfnamefont{X.}~\bibnamefont{Zhang}},
  \bibinfo{author}{\bibfnamefont{L.}~\bibnamefont{Kong}},
  \bibinfo{author}{\bibfnamefont{Q.}~\bibnamefont{Pan}}, \bibnamefont{and}
  \bibinfo{author}{\bibfnamefont{S.}~\bibnamefont{Ge}},
  \bibinfo{journal}{Applied physics letters} \textbf{\bibinfo{volume}{58}},
  \bibinfo{pages}{2042} (\bibinfo{year}{1991}{\natexlab{a}}).

\bibitem[{\citenamefont{Yang et~al.}(1991{\natexlab{b}})\citenamefont{Yang,
  Zhang, Kong, Pan, Ge, Yang, Ding, Zhang, Ye, and Jin}}]{yang.ssc1991}
\bibinfo{author}{\bibfnamefont{Y.}~\bibnamefont{Yang}},
  \bibinfo{author}{\bibfnamefont{X.}~\bibnamefont{Zhang}},
  \bibinfo{author}{\bibfnamefont{L.}~\bibnamefont{Kong}},
  \bibinfo{author}{\bibfnamefont{Q.}~\bibnamefont{Pan}},
  \bibinfo{author}{\bibfnamefont{S.}~\bibnamefont{Ge}},
  \bibinfo{author}{\bibfnamefont{J.}~\bibnamefont{Yang}},
  \bibinfo{author}{\bibfnamefont{Y.}~\bibnamefont{Ding}},
  \bibinfo{author}{\bibfnamefont{B.}~\bibnamefont{Zhang}},
  \bibinfo{author}{\bibfnamefont{C.}~\bibnamefont{Ye}}, \bibnamefont{and}
  \bibinfo{author}{\bibfnamefont{L.}~\bibnamefont{Jin}},
  \bibinfo{journal}{Solid State Communications} \textbf{\bibinfo{volume}{78}},
  \bibinfo{pages}{313 } (\bibinfo{year}{1991}{\natexlab{b}}).

\bibitem[{\citenamefont{Hurley and Coey}(1992)}]{hurley.jpcm1992}
\bibinfo{author}{\bibfnamefont{D.}~\bibnamefont{Hurley}} \bibnamefont{and}
  \bibinfo{author}{\bibfnamefont{J.}~\bibnamefont{Coey}},
  \bibinfo{journal}{Journal of Physics: Condensed Matter}
  \textbf{\bibinfo{volume}{4}}, \bibinfo{pages}{5573} (\bibinfo{year}{1992}).

\bibitem[{\citenamefont{Qi et~al.}(1992)\citenamefont{Qi, Li, and
  Coey}}]{qi.jpcm1992}
\bibinfo{author}{\bibfnamefont{Q.}~\bibnamefont{Qi}},
  \bibinfo{author}{\bibfnamefont{Y.}~\bibnamefont{Li}}, \bibnamefont{and}
  \bibinfo{author}{\bibfnamefont{J.}~\bibnamefont{Coey}},
  \bibinfo{journal}{Journal of Physics: Condensed Matter}
  \textbf{\bibinfo{volume}{4}}, \bibinfo{pages}{8209} (\bibinfo{year}{1992}).

\bibitem[{\citenamefont{Li et~al.}(1993)\citenamefont{Li, Zhou, and
  Morrish}}]{li.jpcm1993}
\bibinfo{author}{\bibfnamefont{Z.}~\bibnamefont{Li}},
  \bibinfo{author}{\bibfnamefont{X.}~\bibnamefont{Zhou}}, \bibnamefont{and}
  \bibinfo{author}{\bibfnamefont{A.}~\bibnamefont{Morrish}},
  \bibinfo{journal}{Journal of Physics: Condensed Matter}
  \textbf{\bibinfo{volume}{5}}, \bibinfo{pages}{3027} (\bibinfo{year}{1993}).

\bibitem[{\citenamefont{Nikitin et~al.}(1999)\citenamefont{Nikitin, Tereshina,
  Verbetsky, and Salamova}}]{nikitin.ijhe1999}
\bibinfo{author}{\bibfnamefont{S.}~\bibnamefont{Nikitin}},
  \bibinfo{author}{\bibfnamefont{I.}~\bibnamefont{Tereshina}},
  \bibinfo{author}{\bibfnamefont{V.}~\bibnamefont{Verbetsky}},
  \bibnamefont{and} \bibinfo{author}{\bibfnamefont{A.}~\bibnamefont{Salamova}},
  \bibinfo{journal}{International Journal of Hydrogen Energy}
  \textbf{\bibinfo{volume}{24}}, \bibinfo{pages}{217} (\bibinfo{year}{1999}).

\bibitem[{\citenamefont{Nikitin et~al.}(2001)\citenamefont{Nikitin, Tereshina,
  Verbetsky, and Salamova}}]{nikitin.jac2001}
\bibinfo{author}{\bibfnamefont{S.}~\bibnamefont{Nikitin}},
  \bibinfo{author}{\bibfnamefont{I.}~\bibnamefont{Tereshina}},
  \bibinfo{author}{\bibfnamefont{V.}~\bibnamefont{Verbetsky}},
  \bibnamefont{and} \bibinfo{author}{\bibfnamefont{A.}~\bibnamefont{Salamova}},
  \bibinfo{journal}{Journal of Alloys and Compounds}
  \textbf{\bibinfo{volume}{316}}, \bibinfo{pages}{46 } (\bibinfo{year}{2001}).

\bibitem[{\citenamefont{Zhang et~al.}(1995)\citenamefont{Zhang, Zhang, Chuang,
  Zhang, Yang, and Du}}]{dan.jpcm1995}
\bibinfo{author}{\bibfnamefont{D.}~\bibnamefont{Zhang}},
  \bibinfo{author}{\bibfnamefont{Z.}~\bibnamefont{Zhang}},
  \bibinfo{author}{\bibfnamefont{Y.}~\bibnamefont{Chuang}},
  \bibinfo{author}{\bibfnamefont{B.}~\bibnamefont{Zhang}},
  \bibinfo{author}{\bibfnamefont{J.}~\bibnamefont{Yang}}, \bibnamefont{and}
  \bibinfo{author}{\bibfnamefont{H.}~\bibnamefont{Du}},
  \bibinfo{journal}{Journal of Physics: Condensed Matter}
  \textbf{\bibinfo{volume}{7}}, \bibinfo{pages}{2587} (\bibinfo{year}{1995}).

\bibitem[{\citenamefont{Coey et~al.}(1993)\citenamefont{Coey, Allan, Minakov,
  and Bugaslavsky}}]{coey.jap1993}
\bibinfo{author}{\bibfnamefont{J.}~\bibnamefont{Coey}},
  \bibinfo{author}{\bibfnamefont{J.}~\bibnamefont{Allan}},
  \bibinfo{author}{\bibfnamefont{A.}~\bibnamefont{Minakov}}, \bibnamefont{and}
  \bibinfo{author}{\bibfnamefont{Y.}~\bibnamefont{Bugaslavsky}},
  \bibinfo{journal}{Journal of Applied Physics} \textbf{\bibinfo{volume}{73}},
  \bibinfo{pages}{5430} (\bibinfo{year}{1993}).

\bibitem[{\citenamefont{Chaboy et~al.}(1995)\citenamefont{Chaboy, Marcelli,
  Bozukov, Baudelet, Dartyge, Fontaine, and Pizzini}}]{chaboy.prb1995}
\bibinfo{author}{\bibfnamefont{J.}~\bibnamefont{Chaboy}},
  \bibinfo{author}{\bibfnamefont{A.}~\bibnamefont{Marcelli}},
  \bibinfo{author}{\bibfnamefont{L.}~\bibnamefont{Bozukov}},
  \bibinfo{author}{\bibfnamefont{F.}~\bibnamefont{Baudelet}},
  \bibinfo{author}{\bibfnamefont{E.}~\bibnamefont{Dartyge}},
  \bibinfo{author}{\bibfnamefont{A.}~\bibnamefont{Fontaine}}, \bibnamefont{and}
  \bibinfo{author}{\bibfnamefont{S.}~\bibnamefont{Pizzini}},
  \bibinfo{journal}{Phys. Rev. B} \textbf{\bibinfo{volume}{51}},
  \bibinfo{pages}{9005} (\bibinfo{year}{1995}).

\bibitem[{\citenamefont{Nordstr\"om et~al.}(1990)\citenamefont{Nordstr\"om,
  Eriksson, Brooks, and Johansson}}]{nordstrom.prb1990}
\bibinfo{author}{\bibfnamefont{L.}~\bibnamefont{Nordstr\"om}},
  \bibinfo{author}{\bibfnamefont{O.}~\bibnamefont{Eriksson}},
  \bibinfo{author}{\bibfnamefont{M.}~\bibnamefont{Brooks}}, \bibnamefont{and}
  \bibinfo{author}{\bibfnamefont{B.}~\bibnamefont{Johansson}},
  \bibinfo{journal}{Phys. Rev. B} \textbf{\bibinfo{volume}{41}},
  \bibinfo{pages}{9111} (\bibinfo{year}{1990}).

\bibitem[{\citenamefont{Eriksson et~al.}(1988)\citenamefont{Eriksson,
  Nordstr\"om, Brooks, and Johansson}}]{eriksson.prl1988}
\bibinfo{author}{\bibfnamefont{O.}~\bibnamefont{Eriksson}},
  \bibinfo{author}{\bibfnamefont{L.}~\bibnamefont{Nordstr\"om}},
  \bibinfo{author}{\bibfnamefont{M.}~\bibnamefont{Brooks}}, \bibnamefont{and}
  \bibinfo{author}{\bibfnamefont{B.}~\bibnamefont{Johansson}},
  \bibinfo{journal}{Phys. Rev. Lett.} \textbf{\bibinfo{volume}{60}},
  \bibinfo{pages}{2523} (\bibinfo{year}{1988}).

\bibitem[{\citenamefont{Alam et~al.}(2013)\citenamefont{Alam, Khan, McCallum,
  and Johnson}}]{alam.apl2013}
\bibinfo{author}{\bibfnamefont{A.}~\bibnamefont{Alam}},
  \bibinfo{author}{\bibfnamefont{M.}~\bibnamefont{Khan}},
  \bibinfo{author}{\bibfnamefont{R.~W.} \bibnamefont{McCallum}},
  \bibnamefont{and} \bibinfo{author}{\bibfnamefont{D.~D.}
  \bibnamefont{Johnson}}, \bibinfo{journal}{Applied Physics Letters}
  \textbf{\bibinfo{volume}{102}}, \bibinfo{eid}{042402} (\bibinfo{year}{2013}).

\bibitem[{\citenamefont{Wang et~al.}(2001{\natexlab{b}})\citenamefont{Wang,
  Shen, Chen, and Wang}}]{wang.jac2001}
\bibinfo{author}{\bibfnamefont{Y.}~\bibnamefont{Wang}},
  \bibinfo{author}{\bibfnamefont{J.}~\bibnamefont{Shen}},
  \bibinfo{author}{\bibfnamefont{N.}~\bibnamefont{Chen}}, \bibnamefont{and}
  \bibinfo{author}{\bibfnamefont{J.}~\bibnamefont{Wang}},
  \bibinfo{journal}{Journal of Alloys and Compounds}
  \textbf{\bibinfo{volume}{319}}, \bibinfo{pages}{62}
  (\bibinfo{year}{2001}{\natexlab{b}}).

\bibitem[{\citenamefont{Ke et~al.}(2015)\citenamefont{Ke, Kukusta, Mccallum,
  and Antropov}}]{ke.intermag2015}
\bibinfo{author}{\bibfnamefont{L.}~\bibnamefont{Ke}},
  \bibinfo{author}{\bibfnamefont{D.}~\bibnamefont{Kukusta}},
  \bibinfo{author}{\bibfnamefont{R.}~\bibnamefont{Mccallum}}, \bibnamefont{and}
  \bibinfo{author}{\bibfnamefont{V.}~\bibnamefont{Antropov}}, in
  \emph{\bibinfo{booktitle}{IEEE Magnetics Conference (INTERMAG)}}
  (\bibinfo{year}{2015}).

\bibitem[{\citenamefont{Li et~al.}(1991)\citenamefont{Li, Zhou, and
  Morrish}}]{li.jap1991}
\bibinfo{author}{\bibfnamefont{Z.}~\bibnamefont{Li}},
  \bibinfo{author}{\bibfnamefont{X.}~\bibnamefont{Zhou}}, \bibnamefont{and}
  \bibinfo{author}{\bibfnamefont{A.}~\bibnamefont{Morrish}},
  \bibinfo{journal}{Journal of Applied Physics} \textbf{\bibinfo{volume}{69}},
  \bibinfo{pages}{5602} (\bibinfo{year}{1991}).

\bibitem[{\citenamefont{Liang et~al.}(1999)\citenamefont{Liang, Huang, Santoro,
  Wang, and Yang}}]{liang.jap1999}
\bibinfo{author}{\bibfnamefont{J.}~\bibnamefont{Liang}},
  \bibinfo{author}{\bibfnamefont{Q.}~\bibnamefont{Huang}},
  \bibinfo{author}{\bibfnamefont{A.}~\bibnamefont{Santoro}},
  \bibinfo{author}{\bibfnamefont{J.}~\bibnamefont{Wang}}, \bibnamefont{and}
  \bibinfo{author}{\bibfnamefont{F.}~\bibnamefont{Yang}},
  \bibinfo{journal}{Journal of Applied Physics} \textbf{\bibinfo{volume}{86}},
  \bibinfo{pages}{2155} (\bibinfo{year}{1999}).

\bibitem[{\citenamefont{Andersen}(1975)}]{andersen.prb1975}
\bibinfo{author}{\bibfnamefont{O.}~\bibnamefont{Andersen}},
  \bibinfo{journal}{Phys. Rev. B} \textbf{\bibinfo{volume}{12}},
  \bibinfo{pages}{3060} (\bibinfo{year}{1975}).

\bibitem[{\citenamefont{Methfessel et~al.}(2000)\citenamefont{Methfessel, van
  Schilfgaarde, and Casali}}]{methfessel.chap2000}
\bibinfo{author}{\bibfnamefont{M.}~\bibnamefont{Methfessel}},
  \bibinfo{author}{\bibfnamefont{M.}~\bibnamefont{van Schilfgaarde}},
  \bibnamefont{and} \bibinfo{author}{\bibfnamefont{R.~A.}
  \bibnamefont{Casali}}, in \emph{\bibinfo{booktitle}{Lecture Notes in
  Physics}}, edited by
  \bibinfo{editor}{\bibfnamefont{H.}~\bibnamefont{Dreysse}}
  (\bibinfo{publisher}{Springer-Verlag, Berlin}, \bibinfo{year}{2000}), vol.
  \bibinfo{volume}{535}.

\bibitem[{\citenamefont{Mackintosh and Andersen}(1980)}]{mackintosh1980}
\bibinfo{author}{\bibfnamefont{A.}~\bibnamefont{Mackintosh}} \bibnamefont{and}
  \bibinfo{author}{\bibfnamefont{O.}~\bibnamefont{Andersen}},
  \emph{\bibinfo{title}{Electrons at the Fermi Surface}}
  (\bibinfo{publisher}{Cambridge University Press},
  \bibinfo{address}{Cambridge, England}, \bibinfo{year}{1980}).

\bibitem[{\citenamefont{Antropov et~al.}(2014)\citenamefont{Antropov, Ke, and
  $\AA$berg}}]{antropov.ssc2014}
\bibinfo{author}{\bibfnamefont{V.}~\bibnamefont{Antropov}},
  \bibinfo{author}{\bibfnamefont{L.}~\bibnamefont{Ke}}, \bibnamefont{and}
  \bibinfo{author}{\bibfnamefont{D.}~\bibnamefont{$\AA$berg}},
  \bibinfo{journal}{Solid State Communications} \textbf{\bibinfo{volume}{194}},
  \bibinfo{pages}{35 } (\bibinfo{year}{2014}).

\bibitem[{\citenamefont{Ke and van Schilfgaarde}(2015)}]{ke.prb2015}
\bibinfo{author}{\bibfnamefont{L.}~\bibnamefont{Ke}} \bibnamefont{and}
  \bibinfo{author}{\bibfnamefont{M.}~\bibnamefont{van Schilfgaarde}},
  \bibinfo{journal}{Phys. Rev. B} \textbf{\bibinfo{volume}{92}},
  \bibinfo{pages}{014423} (\bibinfo{year}{2015}).

\bibitem[{\citenamefont{Ke et~al.}(2013)\citenamefont{Ke, Belashchenko, van
  Schilfgaarde, Kotani, and Antropov}}]{ke.prb2013}
\bibinfo{author}{\bibfnamefont{L.}~\bibnamefont{Ke}},
  \bibinfo{author}{\bibfnamefont{K.}~\bibnamefont{Belashchenko}},
  \bibinfo{author}{\bibfnamefont{M.}~\bibnamefont{van Schilfgaarde}},
  \bibinfo{author}{\bibfnamefont{T.}~\bibnamefont{Kotani}}, \bibnamefont{and}
  \bibinfo{author}{\bibfnamefont{V.}~\bibnamefont{Antropov}},
  \bibinfo{journal}{Phys. Rev. B} \textbf{\bibinfo{volume}{88}},
  \bibinfo{pages}{024404} (\bibinfo{year}{2013}).

\bibitem[{\citenamefont{Ke et~al.}(2012)\citenamefont{Ke, van Schilfgaarde, and
  Antropov}}]{ke.prb2012}
\bibinfo{author}{\bibfnamefont{L.}~\bibnamefont{Ke}},
  \bibinfo{author}{\bibfnamefont{M.}~\bibnamefont{van Schilfgaarde}},
  \bibnamefont{and} \bibinfo{author}{\bibfnamefont{V.}~\bibnamefont{Antropov}},
  \bibinfo{journal}{Phys. Rev. B} \textbf{\bibinfo{volume}{86}},
  \bibinfo{pages}{020402} (\bibinfo{year}{2012}), \bibinfo{note}{$\textbf{Rapid
  communication}$}.

\bibitem[{\citenamefont{von Barth and Hedin}(1972)}]{barth}
\bibinfo{author}{\bibfnamefont{U.}~\bibnamefont{von Barth}} \bibnamefont{and}
  \bibinfo{author}{\bibfnamefont{L.}~\bibnamefont{Hedin}},
  \bibinfo{journal}{Journal of Physics C: Solid State Physics}
  \textbf{\bibinfo{volume}{5}}, \bibinfo{pages}{1629} (\bibinfo{year}{1972}).

\bibitem[{\citenamefont{Haas et~al.}(2009)\citenamefont{Haas, Tran, and
  Blaha}}]{haas.prb2009}
\bibinfo{author}{\bibfnamefont{P.}~\bibnamefont{Haas}},
  \bibinfo{author}{\bibfnamefont{F.}~\bibnamefont{Tran}}, \bibnamefont{and}
  \bibinfo{author}{\bibfnamefont{P.}~\bibnamefont{Blaha}},
  \bibinfo{journal}{Phys. Rev. B} \textbf{\bibinfo{volume}{79}},
  \bibinfo{pages}{085104} (\bibinfo{year}{2009}).

\bibitem[{\citenamefont{Kresse and Hafner}(1993)}]{kresse.prb1993}
\bibinfo{author}{\bibfnamefont{G.}~\bibnamefont{Kresse}} \bibnamefont{and}
  \bibinfo{author}{\bibfnamefont{J.}~\bibnamefont{Hafner}},
  \bibinfo{journal}{Phys. Rev. B} \textbf{\bibinfo{volume}{47}},
  \bibinfo{pages}{558} (\bibinfo{year}{1993}).

\bibitem[{\citenamefont{Kresse and Furthm\"uller}(1996)}]{kresse.prb1996}
\bibinfo{author}{\bibfnamefont{G.}~\bibnamefont{Kresse}} \bibnamefont{and}
  \bibinfo{author}{\bibfnamefont{J.}~\bibnamefont{Furthm\"uller}},
  \bibinfo{journal}{Phys. Rev. B} \textbf{\bibinfo{volume}{54}},
  \bibinfo{pages}{11169} (\bibinfo{year}{1996}).

\bibitem[{\citenamefont{Kresse and Joubert}(1999)}]{kresse.prb1999}
\bibinfo{author}{\bibfnamefont{G.}~\bibnamefont{Kresse}} \bibnamefont{and}
  \bibinfo{author}{\bibfnamefont{D.}~\bibnamefont{Joubert}},
  \bibinfo{journal}{Phys. Rev. B} \textbf{\bibinfo{volume}{59}},
  \bibinfo{pages}{1758} (\bibinfo{year}{1999}).

\bibitem[{\citenamefont{Obbade et~al.}(1997)\citenamefont{Obbade, Fruchart,
  Bououdina, Miraglia, Soubeyroux, and Isnard}}]{obbade.jac1997}
\bibinfo{author}{\bibfnamefont{S.}~\bibnamefont{Obbade}},
  \bibinfo{author}{\bibfnamefont{D.}~\bibnamefont{Fruchart}},
  \bibinfo{author}{\bibfnamefont{M.}~\bibnamefont{Bououdina}},
  \bibinfo{author}{\bibfnamefont{S.}~\bibnamefont{Miraglia}},
  \bibinfo{author}{\bibfnamefont{J.}~\bibnamefont{Soubeyroux}},
  \bibnamefont{and} \bibinfo{author}{\bibfnamefont{O.}~\bibnamefont{Isnard}},
  \bibinfo{journal}{Journal of Alloys and Compounds}
  \textbf{\bibinfo{volume}{253–254}}, \bibinfo{pages}{298 }
  (\bibinfo{year}{1997}).

\bibitem[{\citenamefont{Moze et~al.}(1988)\citenamefont{Moze, Pareti, Solzi,
  and David}}]{moze.ssc1988}
\bibinfo{author}{\bibfnamefont{O.}~\bibnamefont{Moze}},
  \bibinfo{author}{\bibfnamefont{L.}~\bibnamefont{Pareti}},
  \bibinfo{author}{\bibfnamefont{M.}~\bibnamefont{Solzi}}, \bibnamefont{and}
  \bibinfo{author}{\bibfnamefont{W.}~\bibnamefont{David}},
  \bibinfo{journal}{Solid State Communications} \textbf{\bibinfo{volume}{66}},
  \bibinfo{pages}{465 } (\bibinfo{year}{1988}).

\bibitem[{\citenamefont{Goll et~al.}(2014)\citenamefont{Goll, Loeffler, Stein,
  Pflanz, Goeb, Karimi, and Schneider}}]{goll.pss2014}
\bibinfo{author}{\bibfnamefont{D.}~\bibnamefont{Goll}},
  \bibinfo{author}{\bibfnamefont{R.}~\bibnamefont{Loeffler}},
  \bibinfo{author}{\bibfnamefont{R.}~\bibnamefont{Stein}},
  \bibinfo{author}{\bibfnamefont{U.}~\bibnamefont{Pflanz}},
  \bibinfo{author}{\bibfnamefont{S.}~\bibnamefont{Goeb}},
  \bibinfo{author}{\bibfnamefont{R.}~\bibnamefont{Karimi}}, \bibnamefont{and}
  \bibinfo{author}{\bibfnamefont{G.}~\bibnamefont{Schneider}},
  \bibinfo{journal}{physica status solidi (RRL) – Rapid Research Letters}
  \textbf{\bibinfo{volume}{8}}, \bibinfo{pages}{862 } (\bibinfo{year}{2014}).

\bibitem[{\citenamefont{Zhao et~al.}(2014)\citenamefont{Zhao, Nguyen, Zhang,
  Wang, Kramer, Sellmyer, Li, Zhang, Ke, Antropov et~al.}}]{zhao.prl2014}
\bibinfo{author}{\bibfnamefont{X.}~\bibnamefont{Zhao}},
  \bibinfo{author}{\bibfnamefont{M.}~\bibnamefont{Nguyen}},
  \bibinfo{author}{\bibfnamefont{W.}~\bibnamefont{Zhang}},
  \bibinfo{author}{\bibfnamefont{C.}~\bibnamefont{Wang}},
  \bibinfo{author}{\bibfnamefont{M.}~\bibnamefont{Kramer}},
  \bibinfo{author}{\bibfnamefont{D.}~\bibnamefont{Sellmyer}},
  \bibinfo{author}{\bibfnamefont{X.}~\bibnamefont{Li}},
  \bibinfo{author}{\bibfnamefont{F.}~\bibnamefont{Zhang}},
  \bibinfo{author}{\bibfnamefont{L.}~\bibnamefont{Ke}},
  \bibinfo{author}{\bibfnamefont{V.}~\bibnamefont{Antropov}},
  \bibnamefont{et~al.}, \bibinfo{journal}{Phys. Rev. Lett.}
  \textbf{\bibinfo{volume}{112}}, \bibinfo{pages}{045502}
  (\bibinfo{year}{2014}).

\bibitem[{\citenamefont{Trygg et~al.}(1992)\citenamefont{Trygg, Johansson, and
  Brooks}}]{trygg.jmmm1992}
\bibinfo{author}{\bibfnamefont{J.}~\bibnamefont{Trygg}},
  \bibinfo{author}{\bibfnamefont{B.}~\bibnamefont{Johansson}},
  \bibnamefont{and} \bibinfo{author}{\bibfnamefont{M.}~\bibnamefont{Brooks}},
  \bibinfo{journal}{Journal of Magnetism and Magnetic Materials}
  \textbf{\bibinfo{volume}{104}}, \bibinfo{pages}{1447} (\bibinfo{year}{1992}).

\bibitem[{\citenamefont{Hu et~al.}(1989)\citenamefont{Hu, Li, Gavigan, and
  Coey}}]{hu.jpcm1989}
\bibinfo{author}{\bibfnamefont{B.}~\bibnamefont{Hu}},
  \bibinfo{author}{\bibfnamefont{H.}~\bibnamefont{Li}},
  \bibinfo{author}{\bibfnamefont{J.}~\bibnamefont{Gavigan}}, \bibnamefont{and}
  \bibinfo{author}{\bibfnamefont{J.}~\bibnamefont{Coey}},
  \bibinfo{journal}{Journal of Physics: Condensed Matter}
  \textbf{\bibinfo{volume}{1}}, \bibinfo{pages}{755} (\bibinfo{year}{1989}).

\bibitem[{\citenamefont{Thuy et~al.}(1988)\citenamefont{Thuy, Franse, Hong, and
  Hien}}]{thuy.jpc1988}
\bibinfo{author}{\bibfnamefont{N.}~\bibnamefont{Thuy}},
  \bibinfo{author}{\bibfnamefont{J.}~\bibnamefont{Franse}},
  \bibinfo{author}{\bibfnamefont{N.}~\bibnamefont{Hong}}, \bibnamefont{and}
  \bibinfo{author}{\bibfnamefont{T.}~\bibnamefont{Hien}}, \bibinfo{journal}{J.
  Phys. Colloques} \textbf{\bibinfo{volume}{49}}, \bibinfo{pages}{499}
  (\bibinfo{year}{1988}).

\bibitem[{\citenamefont{Tereshina et~al.}(2005)\citenamefont{Tereshina,
  Telegina, Palewski, Skokov, Tereshina, Folcik, and
  Drulis}}]{tereshina.jac2005}
\bibinfo{author}{\bibfnamefont{E.}~\bibnamefont{Tereshina}},
  \bibinfo{author}{\bibfnamefont{I.}~\bibnamefont{Telegina}},
  \bibinfo{author}{\bibfnamefont{T.}~\bibnamefont{Palewski}},
  \bibinfo{author}{\bibfnamefont{K.}~\bibnamefont{Skokov}},
  \bibinfo{author}{\bibfnamefont{I.}~\bibnamefont{Tereshina}},
  \bibinfo{author}{\bibfnamefont{L.}~\bibnamefont{Folcik}}, \bibnamefont{and}
  \bibinfo{author}{\bibfnamefont{H.}~\bibnamefont{Drulis}},
  \bibinfo{journal}{Journal of Alloys and Compounds}
  \textbf{\bibinfo{volume}{404}}, \bibinfo{pages}{208} (\bibinfo{year}{2005}).

\end{thebibliography}
\bigskip 

\end{document}